\documentclass[usenatbib]{emulateapj}
\usepackage{graphicx}
\usepackage[flushleft]{threeparttable}
\usepackage[usenames,dvipsnames,svgnames,table]{xcolor}
\usepackage{hyperref}
\definecolor{darkblue}{rgb}{0.0,0.0,0.3}
\hypersetup{colorlinks,breaklinks,
            linkcolor=darkblue,urlcolor=darkblue,
            anchorcolor=darkblue,citecolor=darkblue}
\usepackage{amsmath,amssymb}
\usepackage{amsmath}

\begin{document}

\def\etal{et al.\ \rm}
\def\ba{\begin{eqnarray}}
\def\ea{\end{eqnarray}}
\def\etal{et al.\ \rm}
\def\Fdw{F_{\rm dw}}
\def\Tex{T_{\rm ex}}
\def\Fdis{F_{\rm dw,dis}}
\def\Fnu{F_\nu}
\def\FJ{F_{J}}
\def\FJE{F_{J,{\rm Edd}}}

\title{Tatooine nurseries: structure and evolution of  
circumbinary protoplanetary disks}

\author{David Vartanyan\altaffilmark{1,2}, Jos\'e A. Garmilla\altaffilmark{1} \& 
Roman R. Rafikov\altaffilmark{1,3}}
\altaffiltext{1}{Department of Astrophysical Sciences, 
Princeton University, Ivy Lane, Princeton, NJ 08540}
\altaffiltext{2}{dvartany@princeton.edu}
\altaffiltext{3}{Institute for Advanced Study, 1 Einstein Drive, Princeton NJ 08540; 
rrr@ias.edu}

%%%%%%%%%%%%%%%%%%%%%%%%%%%%%%%%%%%%%%%%%%%%%%%%%%%%%%%%%%%

\begin{abstract}
Recent discoveries of circumbinary planets by {\it Kepler} mission provide motivation for understanding their birthplaces --- protoplanetary disks around stellar binaries with separations $\lesssim 1$ AU. We explore properties and evolution of such circumbinary disks focusing on modification of their structure caused by tidal coupling to the binary. We develop a set of analytical scaling relations describing viscous evolution of the disk properties, which are verified and calibrated using 1D numerical calculations with realistic inputs. Injection of angular momentum by the central binary suppresses mass accretion onto the binary and causes radial distribution of the viscous angular momentum flux $\FJ$ to be different from that in a standard accretion disk around a single star with no torque at the center. Disks with no mass accretion at the center develop $\FJ$ profile which is flat in radius. Radial profiles of temperature and surface density are also quite different from those in disks around single stars. Damping of the density waves driven by the binary and viscous dissipation dominate heating of the inner disk (within 1-2 AU), pushing the iceline beyond 3-5 AU, depending on disk mass and age. Irradiation by the binary governs disk  thermodynamics beyond $\sim 10$ AU. However, self-shadowing by the hot inner disk may render central illumination irrelevant out to $\sim 20$ AU. Spectral energy distribution of a circumbinary disk exhibits a distinctive bump around 10$\mu$m, which may facilitate identification of such disks around unresolved binaries. Efficient tidal coupling to the disk drives orbital inspiral of the binary and may cause low-mass and relatively compact binaries to merge into a single star within the disk lifetime. We generally find that circumbinary disks present favorable sites for planet formation (despite their wider zone of volatile depletion), in agreement with the statistics of {\it Kepler} circumbinary planets.
\end{abstract}


%%%%%%%%%%%%%%%%%%%%%%%%%%%%%%%%%%%%%%%%%%%%%%%%%%%%%%%%%%%

\section{Introduction.}  
\label{sect:intro}

Theory of planet formation has experienced a wave of recent activity motivated by the large increase in the number of known extrasolar planetary systems. Some of these systems are rather unusual, which is nicely illustrated by the recent discoveries of {\it circumbinary planets} (affectionately termed "Tatooines") around about a dozen of eclipsing binaries by the {\it Kepler} mission \citep{doyle_2011,welsh_2012,orosz_2012a,schwamb_2012}. Central binaries in these systems are composed of main sequence stars with masses around or below $M_\odot$. They have semi-major axes $a_b=0.08-0.2$ AU and some of them are quite eccentric (binary Kepler-34 has eccentriity $e_b=0.52$). Characteristics of planetary orbits in these systems are measured with good accuracy thanks to the exquisite timing precision of the {\it Kepler} satellite, and they are found to be coplanar with their binaries to within a few degrees.

Further, presence of planets orbiting eclipsing post-common envelope binaries has been suspected based on eclipse timing of these systems. A noteworthy example is the evolved binary NN Serpentis \citep{beuermann_2010,marsh_2014} for which it has been argued that if real, the planets would have formed from the material ejected during the common envelope phase \citep{mustill_2013,volschow_2014}.

These discoveries show that planet formation in circumbinary disks is feasible, despite the complications related to the gravitational perturbations induced by the central binary. Recent theoretical efforts have addressed some of these issues
\citep{paardekooper_2012, alexander_2012, meschiari_2012, gong_2013,
foucart_2013, pelupessy_2013, rafikov_2013a, clanton_2013, martin_2013, SR15}. They demonstrated, in particular, that an instrumental part of the theory behind the circumbinary planet genesis is the understanding of the properties and evolution of the circumbinary protoplanetary gaseous disks. Disk properties determine the dynamics of planetary building blocks --- planetesimals --- in many ways. They set not only the efficiency of gas drag but also the precession rates of both planetesimal orbits and the binary itself \citep{rafikov_2013a,SR15}. Even for post-common envelope binaries, it has been contended that a significant fraction of the ejected material forms a circumbinary disk \citep{kashi_2011}.

Circumbinary disks provide a good testing arena for ideas about planet formation, in part because the known systems with circumbinary planets are well constrained, and also because the presence of the binary has important consequences for the evolution of the disk. 
The binary torques are expected to clear out the inner disk region \citep{artymowicz_1994,pelupessy_2013}, significantly reducing the mass supply to the binary (\citet{mcfadyen_2008}, but see \citet{orazio_2013}). Moreover, the
transfer of angular momentum from the binary to the disk changes the structure and dynamical evolution of the disk on large scales. Understanding of the interplay between these complicated processes is necessary to gain insight into the circumbinary planet formation.

The goal of this study is to explore general properties and and main evolutionary features of circumbinary protoplanetary disks around stellar binaries. We focus on understanding the
differences between the characteristics of the circumbinary disks and their more standard counterparts around single stars. In doing this we fully account for the disk thermodynamics influenced both by viscous heating and irradiation by the central binary as well as by tidal interaction between the disk and the binary, through the dissipation of the density waves launched by the binary in the disk. In the course of our study we provide a semi-analytical description for the evolution of disk properties that we test and calibrate using detailed numerical calculations. Our results are then used to assess the implications for circumbinary planet formation, and for the evolution of the central binary itself. 

This paper is organized as follows. After describing our general setup in \S \ref{sect:gen_setup}, we cover the basic of the viscous evolution of circumbinary disks in \S \ref{sect:evolution}. In \S \ref{sect:prop} we describe our treatment of disk thermodynamics. In \S \ref{sect:analytical} we derive a set of analytical scaling relations describing the viscous evolution of the disk properties. We then verify these results numerically in \S \ref{sect:numerical} (our numerical approach is outlined in Appendix \ref{sect:setup}). Spectral differences between the circumbinary and circumstellar disks are described in \S \ref{sect:spectra}. In \S \ref{sect:disc} we discuss the role of different heating sources (\S \ref{sect:heating_terms}), effect of accretion onto the binary (\S \ref{sect:someaccretion}), orbital evolution of the binary due to the tidal coupling with the disk (\S \ref{sect:binangloss}), dead zone (\S \ref{sect:deadzone}) and iceline (\S \ref{sect:iceline}), limitations of our models (\S \ref{sect:limits}) and provide comparison with the existing work on circumbinary disks (\S \ref{sect:others}). We cover the implications for circumbinary planet formation in \S \ref{sect:pl_form} and  conclude with a brief statement of our findings in \S \ref{sect:summ}.

%%%%%%%%%%%%%%%%%%%%%%%%%%%%%%%%%%%%%%%%%%%%%%%%%%%%%%%%%%%
%%%%%%%%%%%%%%%%%%%%%%%%%%%%%%%%%%%%%%%%%%%%%%%%%%%%%%%%%%%

\section{General setup.}  
\label{sect:gen_setup}

%%%%%%%%%%%%%%%%%%%%%%%%%%%%%%%%%%%%%%%%%%%%%%%%%%%%%%%%%%%

We focus on a particular case of a thin disk (we disregard its vertical dimension) coplanar with the binary. The binary has semi-major axis $a_b$, total mass $M_{c}=M_{p}+M_{s}$ ($M_p$ and $M_s$ are the masses of the primary and secondary); the mass ratio of its components is $q\equiv M_s/M_p<1$. 

Our description of the circumbinary disk assumes it to be axisymmetric, with all characteristics --- surface density $\Sigma$, midplane temperature $T$, etc. --- to be functions of radius $r$ only. Thus, we neglect the short-term variability of the disk properties caused by the orbital motion of the binary and focus only on the long-term, time-averaged effect of the binary on the disk. This influence comes mainly in two flavors. 

First, the gravitational potential in which the disk orbits is time-variable, which drives density waves in the disk. Although we do not explore the two-dimensional structure of these perturbations, we do account for this tidal coupling by including the associated angular momentum and energy injection in the disk in our calculations (\S \ref{sect:evolution} and \S \ref{sect:stellar}). But to zeroth order, we model the disk as orbiting in the potential of a point mass $M_c$ centered on the barycenter of the binary. 

Second, illumination of the disk by the binary plays important role in setting its thermal structure (\S \ref{sect:stellar}, \ref{sect:heating_terms}). Orbital motion of the binary leads to periodic variations of the stellar flux impinging on disk surface, the effect of which has been explored in \citet{clanton_2013} and \citet{Quillen}. We neglect this variability of irradiation and study only its time-averaged effect in this work.

%%%%%%%%%%%%%%%%%%%%%%%%%%%%%%%%%%%%%%%%%%%%%%%%%%%%%%%%%%%
%%%%%%%%%%%%%%%%%%%%%%%%%%%%%%%%%%%%%%%%%%%%%%%%%%%%%%%%%%%

\section{Viscous evolution of a circumbinary disk.}  
\label{sect:evolution}

%%%%%%%%%%%%%%%%%%%%%%%%%%%%%%%%%%%%%%%%%%%%%%%%%%%%%%%%%%%

Evolution of the circumbinary disk is driven by both viscous stresses and angular momentum injection by the binary. It is well known that one-dimensional (only in $r$) viscous evolution of the disk surface density $\Sigma$ with external sources of angular momentum can be described by a single diffusion equation \citep{papaloizou_1995}:
\ba
\frac{\partial \Sigma}{\partial t}=-\frac{1}{r}
\frac{\partial}{\partial r}\left[
\left(\frac{d l}
{d r}\right)^{-1}\frac{\partial}{\partial r}
\left(r^3\nu\Sigma\frac{d \Omega}
{d r}\right)+2\frac{\Sigma\Lambda}{\Omega}\right].
\label{eq:evSigma}
\ea
Here $\Omega=\sqrt{GM_{c}/r^{3}}$ is the angular speed and $l=\Omega(r) r^{2}$ is the specific angular momentum in the central binary potential. Following standard approach we model viscosity $\nu$ via the $\alpha$-ansatz 
\citep{shakura_1973}
\ba
\nu=\alpha\frac{c_s^2}{\Omega}=\alpha\frac{k_BT}{\mu\Omega},
\label{eq:alpha_ansatz}
\ea
where $T$ and $c_s$ are the disk (midplane) temperature and sound speed, and $\mu$ is the mean molecular weight of the gas.

The injection of the angular momentum by the binary into the disk can be characterized via the {\it specific angular momentum injection rate} $\Lambda(r)$, with dimensions [cm$^2$ s$^{-2}$] in CGS units. It is defined as the amount of angular momentum transferred by the binary to the disk at radius $r$ per unit time $t$ and unit disk mass. 

In this work we follow \citet{armitage_2002} and adopt
\ba
\Lambda(r) = {\rm sgn}(r-a_b) f \frac{q^{2} G M_{c}}{a_{b}}
\left( \frac{a_{b}}{r-a_{b}} \right)^{4},
\label{eq:Lambda}
\ea
where $f$ is a dimensionless parameter that is chosen as described in Appendix \ref{sect:setup}. This prescription, namely the power law scaling with $|r-a_b|$, is motivated by the analytical calculations of \citet{lin_1979a} and \citet{goldreich_1980}. 
Recent numerical work \citep{mcfadyen_2008, cuadra_2009} shows that torque density in a disk around a high mass ratio binary exhibits considerably more complicated, oscillatory pattern at the cavity edge, rather than following a simple scaling (\ref{eq:Lambda}). The analytical work of \citet{RP12,petrovich_2012} in the low-$q$ limit also suggests that near the gap edge the prescription (\ref{eq:Lambda}) may be incomplete at best. Non-local nature of the density wave damping \citep{GR01,R02} is also not accounted for by equation (\ref{eq:Lambda}). Nevertheless, because of its simplicity and also for reasons outlined in Appendix \ref{sect:setup} we use the prescription (\ref{eq:Lambda}) in this work. It should also be remembered that the last term proportional to $\Lambda$ should be absent in equation (\ref{eq:evSigma}) in the case of a circumstellar disk.

Torque exerted by the binary on the disk clears out a cavity 
in the disk center provided that the mass ratio of the binary 
components is not too small \citep{artymowicz_1994}. For 
$q\sim (0.1-1)$ the cavity radius $r_c$ is about 
twice the binary separation \citep{mcfadyen_2008}, although 
$r_c$ also shows some dependence on the binary eccentricity 
\citep{pelupessy_2013}. At the cavity edge surface density
falls off quite abruptly within a narrow interval of radii.
This is caused by a very steep decay of the torque density
with $r$ \citep{lin_1979a,goldreich_1980,petrovich_2012} --- a point that is 
further discussed in Appendix \ref{sect:setup}. As
a result, small reduction in $r$ leads to large increase of 
$\Lambda$ preventing viscous inflow of material from the disk.
However, it is worth pointing out that non-axisymmetric 
gas motions at the cavity edge (not captured by the 
1D approach based on equation (\ref{eq:evSigma})) do give rise 
to some accretion from the disk into the cavity, albeit at a 
significantly reduced rate $M_b$ than in the case without 
a central binary \citep{mcfadyen_2008}.

\cite{rafikov_2013c} has shown that properties and evolution
of circumbinary disks can in many ways be better understood if instead of $\Sigma$
one uses the {\it viscous angular momentum flux} $F_{J}$ defined as
\ba
\FJ\equiv -2\pi\nu\Sigma r^3\frac{d\Omega}{dr}=3\pi\nu\Sigma l, 
\label{eq:Fnu}
\ea
where the last equality is for a Keplerian disk with 
$\Omega \propto r^{-3/2}$. By definition $\FJ$ is the total 
viscous torque exerted by the inner disk on the outer disk at
a given radius. 

Additionally switching from $r$ to the specific angular momentum $l$, 
the evolution equation (\ref{eq:evSigma}) can be re-written as 
\ba
\frac{\partial}{\partial t} \left( \frac{F_{J}}{D_{J}} \right) =
\frac{\partial}{\partial l} \left[\frac{\partial F_{J}}{\partial l} -
\frac{2F_{J}}{D_{J}} \frac{d \ln l}{d \ln r} \Lambda(l) \right],
\label{eq:evF}
\ea
where
\ba
D_{J}\equiv -\nu r^2\frac{d\Omega}{dr}\frac{dl}{dr}
\label{eq:D_J}
\ea
is the diffusion coefficient. Note that $D_{J}$ depends on $F_{J}$ 
if $\nu$ depends on $\Sigma$.

In terms of the new variables $l$ and $\FJ$, the mass accretion rate 
($\dot M > 0$ for mass inflow) is given by
\ba
\dot{M} \left( l, t \right) = \frac{\partial F_{J}}{\partial l} -
\frac{2F_{J}}{D_{J}} \frac{d \ln l}{d \ln r} \Lambda (l).
\label{eq:mdot}
\ea
This formula shows that the binary torque suppresses mass accretion into the cavity as $\Lambda>0$ for $r>a_b$.

\begin{figure}
\plotone{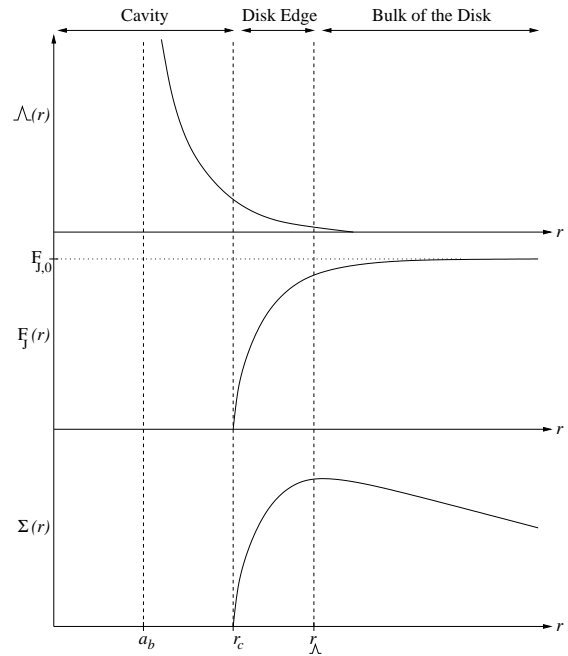}
\caption{Schematic illustration of the behavior of the disk surface density $\Sigma$, angular momentum flux $\FJ$ and specific torque density $\Lambda$ in different parts of the circumbinary disk. Vertical dashed lines show the binary semi-major axis $a_b$, radius of the inner cavity $r_c$ and the radius $r_\Lambda$ beyond which the binary torque density can be neglected. Key disk regions discussed in the text --- cavity ($r\lesssim r_c$), disk edge ($r_c\lesssim r\lesssim r_\Lambda$), and bulk of the disk ($r\gtrsim r_\Lambda$) --- are indicated. The case of non-accreting binary ($\dot M_b\approx 0$) is illustrated, so that $\FJ$ is radially constant in the bulk of the disk.}
\label{fig:sketch}
\end{figure}

%%%%%%%%%%%%%%%%%%%%%%%%%%%%%%%%%%%%%%%%%%%%%%%%%%%%%%%%%%%

\subsection{Disk evolution away from the cavity edge.}  
\label{sect:far_regime}

The aforementioned steep dependence of $\Lambda$ on $r$ implies that the binary torque becomes insignificant for the disk evolution outside some radius $r_\Lambda$, which is not very different from the cavity radius $r_c$ --- radius where $\Sigma$ (and $\FJ$) becomes  small. Thus, it is convenient to separate the disk into two parts, as illustrated in Figure \ref{fig:sketch}. One is the {\it edge region} ($r_c\lesssim r\lesssim r_\Lambda$) in which $\Lambda$ is large and where essentially all of the angular momentum and energy carried by the binary-induced density waves get dissipated. Second is the {\it bulk of the disk} in which $\Lambda$ is negligible and the surface density (and $\FJ$) evolution is determined solely by viscous stresses. 

The utility of switching to $\FJ$ and $l$ variables can be best 
appreciated in the bulk of the disk, $r\gtrsim r_\Lambda$, outside 
the cavity edge region. As $\Lambda\to 0$ there, the equation 
(\ref{eq:evF}) transforms to a particularly simple form \citep{lynden-bell_1974,filipov_1984,lyubarskij_1987,rafikov_2013c}
\ba
\frac{\partial}{\partial t} \left( \frac{F_{J}}{D_{J}} \right) =
\frac{\partial^2\FJ}{\partial l^2}.
\label{eq:evF_simple}
\ea
Also, $\dot{M} \left( l, t \right)= \partial F_{J}/\partial l$ in this case, according to equation (\ref{eq:mdot}).

Equation (\ref{eq:evF_simple}) must be supplemented by a boundary condition (BC) at the inner boundary of the main disk region, i.e. at $r=r_\Lambda$, which characterizes the suppression of the mass inflow by the binary torque. We express this condition via the constraint on the mass accretion rate of gas from the disk into the cavity $\dot M_b$. We assume $\dot M_b$ to be set by the processes happening in the edge region, at $r\lesssim r_\Lambda$ \citep{mcfadyen_2008}. Its value is set by the presice form of $\Lambda(r)$ and can be non-zero in general.

Given the small radial extent of the edge region, it contains rather small amount of mass and one can safely assume that $\dot M$ at the inner boundary of the bulk of the disk, at $r_\Lambda$, is just equal to $\do M_b$. This means that a proper inner BC for equation
(\ref{eq:evF_simple}) is
\ba
\frac{\partial \FJ}{\partial l}\Big|_{l(r_\Lambda)}=\dot M_b.
\label{eq:inn_BC}
\ea
When studying global disk evolution at $r\gg r_\Lambda$, one can simply assume this inner BC to be imposed at $r\to 0$ ($l\to 0$).

In this work we will mainly work with the no-inflow BC, such that $\dot M_b=0$.

%%%%%%%%%%%%%%%%%%%%%%%%%%%%%%%%%%%%%%%%%%%%%%%%%%%%%%%%%%%

\subsection{(Quasi-)steady state circumbinary disks.}  
\label{sect:steady}

%%%%%%%%%%%%%%%%%%%%%%%%%%%%%%%%%%%%%%%%%%%%%%%%%%%%%%%%%%%

We now use the framework outlined above to understand the steady state structure to which a circumbinary disk tends to converge as a result of its viscous evolution. Setting the left-hand side of equation (\ref{eq:evF_simple}) to zero one immediately obtains a steady-state solution for $\FJ$ in a simple form \citep{rafikov_2013c}
\ba
\FJ(l)=F_{J,0}+F_{J,1}l=F_{J,0}+\dot M l,
\label{eq:lin_sol}
\ea
where $\dot M=\partial F_{J}/\partial l=F_{J,1}$ is constant. This is different from a standard constant $\dot M$ accretion disk solution for which $\dot M=3\pi\nu\Sigma$ and it follows from equation (\ref{eq:Fnu}) that 
\ba
\FJ=\dot Ml=\dot M\Omega r^2.
\label{eq:const_Mdot}
\ea
This expression reduces to solution (\ref{eq:lin_sol}) only when $F_{J,0}=0$ and the angular momentum flux at the disk center is zero (i.e. no injection of the angular momentum at the center by e.g. binary torques). 
In a circumbinary disk $F_{J,0}\neq 0$ because binary torques inject non-zero angular momentum at the center, as $l\to 0$. As a result, disk structure differs from that of a standard constant $\dot M$ disk, even though in both cases $\dot M$ is radially constant. This is an important distinction that is not always appreciated. 

Even when $\dot M=0$ circumbinary disk has non-zero $\Sigma$ and dissipation rate, simply because $F_{J,0}\neq 0$. In this limit of non-accreting binary ($\dot M_b\approx 0$) $\FJ$ is {\it radially constant}, which is illustrated in Figure \ref{fig:sketch}. This is a situation that we will mainly encounter in this work, see \S \ref{sect:FJ}. 

All these statements also apply to the {\it quasi-stationary} 
disk evolution, in regions where the viscous time $r^2/\nu$ is 
shorter than the global evolution timescale of the disk. 
In particular, if the disk is supplied with mass from the {\it outside} 
then a solution (\ref{eq:lin_sol}) will develop in its {\it inner} parts,
because the viscous evolution time is short there. This solution 
will have its $F_{J,0}$ and $\dot M$ slowly changing in time on
the global disk evolution timescale $t_\nu$. However, $\FJ$ will 
still exhibit roughly linear dependence on $l$ out to the radius, 
at which viscous timescale $t_\nu$ is of order the disk lifetime 
\citep{rafikov_2013c}. 

A remarkable property of the steady-state solution 
(\ref{eq:lin_sol}) is that the behavior of $\FJ$ is completely independent of the 
details of physical processes operating in the disk --- origin of 
viscous stresses, thermal state of the disk, and so on \citep{rafikov_2013c}. Given 
this solution one can trivially obtain the radial scalings of 
the disk properties by solving simple algebraic equations 
as described in the next section. This justifies the 
introduction of $\FJ$-$l$ variables and makes interpretation of 
our results particularly transparent. Previous treatments of 
the circumbinary disk evolution in terms of $\Sigma$ and $r$ 
\citep{ivanov_1999} offered less straightforward interpretation.

%%%%%%%%%%%%%%%%%%%%%%%%%%%%%%%%%%%%%%%%%%%%%%%%%%%%%%%%%%%
%%%%%%%%%%%%%%%%%%%%%%%%%%%%%%%%%%%%%%%%%%%%%%%%%%%%%%%%%%%

\section{Circumbinary disks around stellar binaries.}  
\label{sect:prop}

%%%%%%%%%%%%%%%%%%%%%%%%%%%%%%%%%%%%%%%%%%%%%%%%%%%%%%%%%%%

Discussion in the previous section makes it straightforward that the structure of the circumbinary disk is best parametrized via the angular momentum flux $\FJ$. This is in contrast to the conventional accretion disks without angular momentum sources, the properties of which are specified via the mass accretion rate $\dot M$. In the circumbinary case such parametrization would be ambiguous, which is easy to understand by looking at the steady state solution (\ref{eq:lin_sol}): $\FJ$ can vary broadly for a given $\dot M$ depending on the value of $F_{J,0}$.

To compute detailed properties of circumbinary disks for a given angular momentum flux $F_J$ at some specific radius $r$ one needs to specify the thermal structure of the disk, since the viscosity depends on the disk temperature, see equation (\ref{eq:alpha_ansatz}). Following standard approach, in \S \ref{sect:stellar} we calculate midplane disk temperature $T$ by considering vertical energy transport in the disk and relating $T$ to $\Sigma$ and one other disk characteristic. This is a well-known exercise in the case of a standard, constant $\dot M$ disk, in which $T$ is related to $\Sigma$ and $\dot M$, allowing one to infer radial profiles of both $\Sigma$ and $T$. 

By using $\FJ$ rather than $\dot M$ as a global disk characteristic, \citet{rafikov_2013c} computed radial behavior of various disk properties as a function of $\FJ$ in hot and luminous circumbinary disks around supermassive black hole binaries. In this work we extend this approach to circumbinary disks around stellar binaries.

%%%%%%%%%%%%%%%%%%%%%%%%%%%%%%%%%%%%%%%%%%%%%%%%%%%%%%%%%%%

\subsection{Disk thermodynamics: heating sources}  
\label{sect:stellar}

%%%%%%%%%%%%%%%%%%%%%%%%%%%%%%%%%%%%%%%%%%%%%%%%%%%%%%%%%%%

Circumbinary disk around a stellar binary receives energy from three main sources. First, there is viscous dissipation within the disk at the rate $d\dot E_v/dr$ per unit time and radius given by 
\ba
\frac{d\dot E_v}{dr}=-F_J\frac{d\Omega}{dr}=
\frac{3}{2} \frac{\FJ \Omega}{r}.
\label{eq:dE_v}
\ea
The associated one-sided energy flux escaping from a disk surface 
is 
\ba
{\cal F}_v=\frac{1}{4\pi r}\frac{d\dot E_v}{dr}=
\frac{3}{8\pi} \frac{\FJ \Omega}{r^2},
\label{eq:flux_nu}
\ea
and is independent of the details of the disk structure as long as 
the value of $\FJ$ is specified. In the limit of a standard
constant $\dot M$ disk with $\FJ$ given by equation 
(\ref{eq:const_Mdot}) one finds the familiar expression 
${\cal F}_v=(3/8\pi)\dot M \Omega^2$ \citep{shakura_1973}. 

Second, the disk is also illuminated by the central binary,
which irradiates its surface at a grazing incidence angle $\zeta\ll 1$.
\cite{chiang_1997} have demonstrated that centrally irradiated 
disks develop a distinct two-layer structure: an outer 
``superheated layer'' intercepts direct stellar radiation
and re-radiates one half of it towards the inner, midplane region,
which comprises most of the disk mass. The irradiation energy 
flux received by a disk surface is thus\footnote{This expression 
assumes that central cavity in the disk allows an unobscured view 
of full stellar disks of both stars from any point on the disk 
surface.}
\ba
{\cal F}_{\rm irr}=\frac{1}{2}\frac{L_c}{4\pi r^2}\zeta,
\label{eq:Firr}
\ea
where $L_c$ is the combined luminosity of the central stars.
This expression ignores the time-dependence of irradiation
caused by the orbital motion of the central binary 
\citep{clanton_2013}.
 
If stellar radiation is intercepted far from the binary
at a height $\eta h(r)$ above the disk midplane
then the incidence angle $\zeta$ can be approximated as
\ba
\zeta\approx r\frac{d}{dr}\left(\frac{\eta h}{r}\right).
\label{eq:zeta}
\ea
Typically $\eta\approx 3-4$, and varies with $r$ very weakly
\citep{chiang_1997}. 

There is also a third energy source --- shock damping of the density 
waves excited by the binary potential, which effect the angular 
momentum exchange between the 
binary and the disk. The one-sided contribution of this {\it tidal} 
heating per unit area is
\begin{equation}
{\cal F}_{\rm tid} = \frac{1}{2}\left(\Omega_b - \Omega\right) \Lambda \Sigma\,,
\label{eq:ftid}
\end{equation}
where $\Omega_b$ is the angular frequency of the binary and the torque $\Lambda$ due the binary is given by equation (\ref{eq:Lambda}).

%%%%%%%%%%%%%%%%%%%%%%%%%%%%%%%%%%%%%%%%%%%%%%%%%%%%%%%%%%%

\subsection{Disk thermodynamics: radiation transport}  
\label{sect:transport}

%%%%%%%%%%%%%%%%%%%%%%%%%%%%%%%%%%%%%%%%%%%%%%%%%%%%%%%%%%%

We characterize the thermal state of the disk by its midplane 
temperature $T$ and take it to represent the characteristic 
temperature across the whole vertical span of the disk, which 
is of course just an approximation. To compute $T$ given the energy
fluxes ${\cal F}_v$, ${\cal F}_{\rm irr}$ and ${\cal F}_{\rm tid}$ 
we assume that energy is transported in the vertical direction 
by radiation and characterize transport by the midplane-to-surface
optical depth $\tau\equiv (1/2)\Sigma\kappa$. Here $\kappa$ is the 
opacity, which in general is a function of $T$ and (midplane) 
gas density $\rho$, which can be easily related to $\Sigma$ or 
$\FJ$. 

In the optically thin limit $\tau\ll 1$ Kirchhoff's law and 
energy balance suggest that
\ba
\tau\sigma T^4\approx {\cal F}_v+{\cal F}_{\rm tid}+\tau{\cal F}_{\rm irr}.
\label{eq:T_thin}
\ea
Note that technically the values of $\tau$ in the left and right 
hand sides of this relation are different: the former is for the 
disk opacity at temperature $T$, while the latter characterizes 
optical depth of the disk to the radiation of super-heated dust
grains at the disk surface \citep{chiang_1997}. The two are
somewhat different (disk surface is hotter) but for simplicity 
we will neglect this distinction when calculating the midplane 
temperature. 

In the optically thick case $\tau\gg 1$ standard solution for
radiation transfer in a plane-parallel atmosphere 
\citep{calvet_1991} yields
\ba
\sigma T^4\approx \frac{3}{8}\tau\left({\cal F}_v+{\cal F}_{\rm tid}\right) +
{\cal F}_{\rm irr}.
\label{eq:T_thick}
\ea
This expression assumes that viscous energy release occurs close 
to midplane. Note that in equations (\ref{eq:T_thin}) and 
(\ref{eq:T_thick}) we do not differentiate between the Planck 
and Rosseland mean opacities.

To cover both opacity limits (and interpolate in the $\tau\sim 1$ 
regime) we use the following expression for $T$, which provides 
smooth transition between the two regimes:
\ba
\sigma T^4 & = & f(\tau)\left({\cal F}_v+{\cal F}_{\rm tid}\right)+
{\cal F}_{\rm irr},
\label{eq:gen_T}\\
%%%%%%%%
f(\tau)& \approx & \frac{3}{8}\tau+\tau^{-1}\,.
\label{eq:gen_tau}
\ea
It is clear that in the limits $\tau\ll 1$ and $\tau\gg 1$ this 
expression reduces to equations (\ref{eq:T_thin}) and 
(\ref{eq:T_thick}) respectively. If vertical energy transport is 
affected not by radiation but by 
convection then the relation (\ref{eq:gen_T}) remains the same 
but the form of $f(\tau)$ changes in the optically thick case  
\citep{rafikov_2007}.

Equations (\ref{eq:alpha_ansatz}), (\ref{eq:Fnu}), 
(\ref{eq:gen_T}) form a closed set of 
equations uniquely determining $\Sigma$ and $T$ for given $\FJ$
and $r$. This fully specifies disk properties, including circumbinary
disks which are not in steady state.

%%%%%%%%%%%%%%%%%%%%%%%%%%%%%%%%%%%%%%%%%%%%%%%%%%%%%%%%%%%

\subsection{Irradiation-dominated disk regions.}  
\label{sect:passive}

We will now describe the disk structure in a particular important limit when the midplane temperature is determined primarily by stellar irradiation, i.e.\,when 
${\cal F}_{\rm irr}\gg f(\tau)\left({\cal F}_{v}+{\cal F}_{\rm tid}\right)$, see equation (\ref{eq:gen_T}). As we will see later, this limit is naturally realized in the outer, cold parts of the circumbinary disk. Calculation of the temperature profile in this regime is essentially identical to that in the single star case, but we show it for completeness as the result is then used in \S \ref{sect:analytical}.

In the irradiation-dominated regime the midplane temperature is
\ba
T(r)\approx \left(\frac{{\cal F}_{\rm irr}}{\sigma}\right)^{1/4}=
\left[\frac{\zeta(r)L_c}{8\pi\sigma}\right]^{1/4}r^{-1/2}.
\label{eq:Tirr_gen}
\ea
Note that the incidence angle $\zeta$ is itself a function of $T$, 
according to equation (\ref{eq:zeta}). Plugging the expression (\ref{eq:zeta}) for 
$\zeta$ into equation (\ref{eq:Tirr_gen}) and solving it, we find 
\citep{chiang_1997,rafikov_2006}
\ba
T(r)&=&\left[\left(\frac{\eta}{7}\frac{L_c}{4\pi\sigma}\right)^{2}
\frac{k/\mu}{(GM_c)}\right]^{1/7}r^{-3/7}
\label{eq:T_irr}
\\
&\approx &
120~\mbox{K}\left[\frac{\eta_3^2 L_{c,1}^2}
{\mu_2 M_{c,1}}\right]^{1/7}r_1^{-3/7}.
\nonumber
\ea
In the numerical estimate we use a 
shorthand notation $r_1\equiv r/$AU, $\eta_3\equiv\eta/3$, 
$L_{c,1}\equiv L_c/L_\odot$, $M_{c,1}=M_c/M_\odot$, and $\mu$
is normalized by the H$_2$ molecular weight. The temperature 
distribution in (\ref{eq:T_irr}) is independent of $\FJ$. 

The aspect ratio of an externally irradiated disk is given by 
\ba
\frac{h}{r}&=&\left[\frac{\eta}{7}
\left(\frac{k}{\mu}\right)^4\frac{L_c}{4\pi\sigma (GM_c)^4}
\right]^{1/7}r^{2/7}
\label{eq:hr_irr}
\\
&\approx &
0.024\left[\frac{\eta_3 L_{c,1}}
{\mu_2^4 M_{c,1}^4}\right]^{1/7}r_1^{2/7},
\nonumber
\ea
so that the angle at which stellar radiation impinges on the disk surface is 
\ba
\zeta\approx (2/7)(h/r). 
\label{eq:zeta_irr}
\ea
From equations (\ref{eq:alpha_ansatz}) and (\ref{eq:Fnu}) we also find
\ba
\Sigma(r)&=&\frac{\FJ}{3\pi\alpha}
\left[\left(\frac{\eta}{7}\frac{L_c}{4\pi\sigma}\right)^{-2}
\frac{(GM_c)}{(k/\mu)^8}\right]^{1/7}r^{-11/7}
\label{eq:surf_irr}
\\
&\approx &
10^3~\mbox{g cm}^{-2}\frac{F_{J,38}}{\alpha_{-2}}
\left[\frac{\mu_2^8 M_{c,1}}{\eta_3^2 L_{c,1}^2}
\right]^{1/7}r_1^{-11/7},
\nonumber
\ea
where $\alpha_{-2}\equiv \alpha/10^{-2}$ and 
$F_{J,38}\equiv \FJ/(10^{38}\mbox{erg})$. The characteristic value 
of $\FJ$ used here is motivated in \S \ref{sect:F_J}, see 
equation (\ref{eq:FJ_irr}).

Note that within the framework of our approximations (i.e. not 
differentiating between the opacity for the radiation of the disk 
midplane and the superheated outer layer) the properties 
of passive regions of the disk are independent of the opacity 
behavior and optical depth.

%%%%%%%%%%%%%%%%%%%%%%%%%%%%%%%%%%%%%%%%%%%%%%%%%%%%%%%%%%%
%%%%%%%%%%%%%%%%%%%%%%%%%%%%%%%%%%%%%%%%%%%%%%%%%%%%%%%%%%%

\section{Circumbinary disk evolution: analytical picture}  
\label{sect:analytical}

%%%%%%%%%%%%%%%%%%%%%%%%%%%%%%%%%%%%%%%%%%%%%%%%%%%%%%%%%%%

Previously obtained results allow us to characterize structure and evolution of a  circumbinary disk. We assume that most of the disk mass is initially deposited at large radii (tens of AU) by the collapse of a centrifugally supported envelope. We will also focus on a particular case of a disk with no or weak mass inflow into the cavity, so that $\dot M_b\approx 0$. 

After initial mass deposition at some radius $r_0$ the disk will evolve under the action of viscous stresses. Some of the mass will spread inwards towards the binary until it reaches semi-major axis $\sim (2-3)a_b$. At this point the binary torque will become strong enough at imparting angular momentum to the disk fluid that the gas inflow will be stopped (or at least strongly suppressed) and a cavity will form at the disk center. At the same time most of the mass remains in the outer disk (see \S \ref{sect:F_J}), which spreads out viscously. As a result, the outer radius of the disk will grow in time. Provided that $\dot M_b\approx 0$ the mass of the disk will be conserved during this evolution while its angular momentum will increase because of the binary torque. This distinguishes circumbinary disks form the conventional circumstellar disks \citep{shakura_1973}, which evolve preserving their total angular momentum (in the absence of external torques) but losing mass to accretion onto the central object.

Characteristic time on which viscous evolution occurs and $\FJ$, $\Sigma$ and other disk properties change at some radius $r$ is the viscous time $t_\nu(r)=r^2/\nu=\alpha^{-1}\Omega r^2(\mu/k_B T(r))$. Global disk evolution is set by $t_\nu$ at the radius where most of the mass is concentrated, i.e. in the outer disk. This is the region where the midplane temperature is determined primarily by central irradiation, so that we can use the results of \S \ref{sect:passive} to characterize disk temperature behavior. Using equations (\ref{eq:alpha_ansatz}) and (\ref{eq:T_irr}) we find in this case 
\ba
t_\nu(r) & = & \alpha^{-1}\left[\frac{7}{\eta}
\left(\frac{\mu}{k}\right)^4\frac{4\pi\sigma}{L_c}(GM_c)^{9/4}
\right]^{2/7}r^{13/14}
\label{eq:t_nu_0}\\
%%%%%%%%%%
& \approx & 4.4\times 10^5~\mbox{yr}
\left[\frac{\mu_2^4 M_{c,1}^{9/4}}{\eta_3 L_{c,1}}\right]^{2/7}
\left(\frac{r}{20~\mbox{AU}}\right)^{13/14}.
\nonumber
\ea

This formula demonstrates that $t_\nu$ goes down as $r$ decreases, implying that the inner regions of the disk will tend to evolve faster and attain a quasi-steady state described in \ref{sect:steady}. Then, because of our assumption $\dot M_b=0$ and according to the reasoning presented \S \ref{sect:steady} the disk will tend to converge to $\FJ(r)=\,$const state, as confirmed by our numerical calculations in \S \ref{sect:FJ}.

%%%%%%%%%%%%%%%%%%%%%%%%%%%%%%%%%%%%%%%%%%%%%%%%%%%%%%%%%%%%%%%%

\subsection{Disk expansion}  
\label{sect:expansion}

We now take a closer look at how the disk expands in the long run, after the age of the system $t$ exceeds the viscous time $t_\nu(r_0)$ at the initial mass deposition radius $r_0$. We characterize disk expansion by the {\it radius of influence} $r_{\rm infl}(t)$, which is the radius at which viscous timescale equals the age of the system $t$. At late times $r_{\rm infl}$ can be obtained by inverting the relation (\ref{eq:t_nu_0}) so that
\ba
r_{\rm infl}(t) &=&
\left[\frac{\eta}{7}\frac{L_c}{4\pi\sigma}
\left(\frac{k}{\mu}\right)^4\right]^{4/13}
\frac{\left(\alpha t\right)^{14/13}}{\left(GM_c\right)^{9/13}}
\label{eq:rinfl}\\
%%%%%%%
& \approx & 160~\mbox{AU}
\left[\frac{\alpha_{-2}^{14}L_{c,1}^4\eta_3^4}
{\mu_2^{16}M_{c,1}^9}\right]^{1/13}
\left(\frac{t}{3~\mbox{Myr}}\right)^{14/13}.
\ea
This relation is expected to be accurate only for $t\gg t_\nu(r_0)$. Note that it predicts $r_{\rm infl}$ to be independent of the disk mass (or $\FJ$), which is a consequence of our natural assumption that irradiation fully determines disk temperature in the outer regions.

%%%%%%%%%%%%%%%%%%%%%%%%%%%%%%%%%%%%%%%%%%%%%%%%%%%%%%%%%%%%%%%%

\subsection{Characteristic value of $F_J$.}  
\label{sect:F_J}

We now motivate the characteristic value of $\FJ=10^{38}$ ergs that has been adopted in our equation (\ref{eq:surf_irr}) for an externally irradiated disk around a stellar mass binary. We do this by relating $F_J$ to the disk mass $M_d(r)$ enclosed 
within some radius $r$. Using definition (\ref{eq:Fnu}) and 
the conventional $\alpha$-prescription for viscosity $\nu$, one 
finds $\Sigma=(3\pi\alpha r^2 c_s^2)^{-1}\FJ$. Integrating 
this over the radial span of the disk with $2\pi r$ weighting 
one finds
\ba
M_d(r)\approx \frac{2}{3\alpha}\int\limits^{r}\frac{\FJ(r^\prime)}{c_s^2(r^\prime)}
\frac{dr^\prime}{r^\prime}.
\label{eq:M_d}
\ea
Note that we dropped the lower integration limit in this 
expression, which is $\sim r_\Lambda$. This is justified 
for any disk with $r\gg r_\Lambda$ in which the outer regions 
contain most of the mass. As equation (\ref{eq:M_d}) shows,  in a constant $\FJ$ disk this is a direct consequence of $T$ falling 
with radius.

Equation (\ref{eq:M_d}) allows us to relate $\FJ$ in a constant-$\FJ$ disk to the total disk mass $M_d$ and its outer radius $r$. Beyond $~10$ AU we expect $c_s$ to be determined by external irradiation, so substituting $T(r)$ from (\ref{eq:T_irr}) into (\ref{eq:M_d}) and performing the integral we express $\FJ$ as
\ba
\FJ &=& \frac{9}{14}\frac{M_d}{r_d^{3/7}}\alpha
\left[\left(\frac{\eta}{7}\frac{L_c}{4\pi\sigma}\right)^{2}
\frac{(k/\mu)^8}{(GM_c)}\right]^{1/7}
\label{eq:FJ_irr}
\\
&\approx &
10^{38}\mbox{erg}~M_{d,-2}\alpha_{-2}
\left[\frac{\eta_3^2L_{c,1}^2}{\mu_{2}^8 M_{c,1}}
\left(\frac{50\mbox{AU}}{r}\right)^3\right]^{1/7},
\nonumber
\ea
where $M_{d,-2}\equiv M_d/(10^{-2}M_\odot)$. This estimate justifies the characteristic value of $F_J=10^{38}$ ergs used in equation (\ref{eq:surf_irr}) for a disk with a typical
mass $10^{-2}M_\odot$ and size $r=50$ AU.

%%%%%%%%%%%%%%%%%%%%%%%%%%%%%%%%%%%%%%%%%%%%%%%%%%%%%%%%%%%%%%%%

\subsection{Evolution of $F_J$.}  
\label{sect:F_J_evol}

We can now predict the late time (for $t\gtrsim t_\nu(r_0)$) behavior of $\FJ$ in a constant $\FJ$ disk. As $M_d$ remains constant because of our assumption $\dot M_b\approx 0$, while the disk size increases due to viscous spreading, equation (\ref{eq:FJ_irr}) implies that $\FJ$ will go down with time. Approximating the disk size at a given moment of time $t$ with $r_{\rm infl}(t)$, we substitute the expression (\ref{eq:rinfl}) for $r$ in equation (\ref{eq:FJ_irr}) to find
\ba
F_J(t) &=& M_d
\left[\frac{\eta}{7}\frac{L_c(GM_c)}{4\pi\sigma}
\left(\frac{k}{\mu}\right)^4\right]^{2/13}
\frac{\alpha^{7/13}}{t^{6/13}}
\label{eq:FJ_via_M_d}\\
%%%%%%%%%%
&\approx & 1.1 \times 10^{38}~\mbox{ergs}~
M_{d,-2}\alpha_{-2}^{7/13}
\nonumber\\
%%%%%%%%%%
&\times &\left[\frac{M_{c,1}L_{c,1}\eta_3}
{\mu_2^4}\right]^{2/13}
\left(\frac{t}{3~\mbox{Myr}}\right)^{-6/13}.
\ea
This formula determines the scalings of the characteristic $\FJ$ in a constant-$\FJ$ disk with the disk mass $M_d$ and the age of the system $t$, which will be subsequently verified in \S \ref{sect:FJ}. It can be used to determine the evolution of $\Sigma(r,t)$ in the outer disk dominated by central irradiation with the aid of equation (\ref{eq:surf_irr}).

Note that equation (\ref{eq:FJ_via_M_d}) does not contain any details of the opacity behavior in the outer disk. This is a direct consequence of our simplified treatment of the radiation transfer, in which we took the dust opacity to be the roughly the same for both the midplane temperature $T$ and the temperature of the superheated surface layers of the disk (recognizing this distinction would would give rise to only small difference with our results).

%%%%%%%%%%%%%%%%%%%%%%%%%%%%%%%%%%%%%%%%%%%%%%%%%%%%%%%%%%%%%%%%
%%%%%%%%%%%%%%%%%%%%%%%%%%%%%%%%%%%%%%%%%%%%%%%%%%%%%%%%%%%%%%%%

\section{Circumbinary disk evolution: numerical results.}
\label{sect:numerical}

%%%%%%%%%%%%%%%%%%%%%%%%%%%%%%%%%%%%%%%%%%%%%%%%%%%%%%%%%%%%%%%%

Now we describe the results of our 1D numerical calculations of the circumbinary disk evolution based on equation (\ref{eq:evSigma}). In doing this we provide close comparison with the analytical predictions outlined in \S \ref{sect:analytical}.\,To better illustrate the differences between the circumbinary and standard constant $\dot M$ circumstellar disks, we also performed evolutionary calculations for disks orbiting a single star with the same total mass $M_c$ and {\it no torque injection} at the center. Our numerical approach and parameters of our simulations are described in Appendix \ref{sect:setup}, which should be referred for details.

%%%%%%%%%%%%%%%%%%%%%%%%%%%%%%%%%%%%%%%%%%%%%%%%%%%%%%%%%%%

\subsection{Viscous expansion of the disk}  
\label{sect:exp}

%%%%%%%%%%%%%%%%%%%%%%%%
\begin{figure}
\plotone{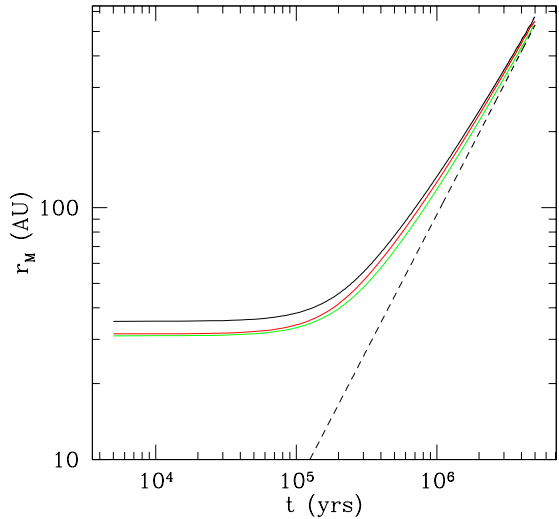}
\caption{Semianalytic fit to the evolution of $r_{\rm infl}(t)$. Solid lines are the results of the numerical calculations, and the dashed line is $r_M(t)$ given by equation (\ref{eq:r_M}) and based on analytic theory (equation (\ref{eq:rinfl})), which fits numerical results at late times. Different colors correspond to different disk masses: black $M_d=0.1 M_c$, red $M_d=0.05 M_c$, and green $M_d=0.01 M_c$.}
\label{fig:r_M}
\end{figure}
%%%%%%%%%%%%%%%%%%%%%%%%

First, we explore the viscous expansion of the disk seen in our simulations. As a proxy for the outer radius of the disk we use the radius $r_M$ at which the ``effective disk mass'' $\Sigma(r)r^2$ attains its maximum value. Figure \ref{fig:r_M} shows a plot of $r_M(t)$ obtained from our simulations for different disk masses. 

At early times, for $t\lesssim t_\nu(r_0)$, one naturally has $r_M\approx r_0$ reflecting the radius of initial mass deposition. Later on, for $t\gtrsim t_\nu(r_0)$, the disk viscously expands and $r_M(t)$ grows. By fitting our numerical results we find that at late times 
\ba
r_M(t) \approx 1.5 r_{\rm infl}(t),
\label{eq:r_M}
\ea
where $r_{\rm infl}(t)$ is given by equation (\ref{eq:rinfl}).

Note that late-time $r_M(t)$ is only weakly dependent of $M_d$ in our simulations. This is in agreement with our analytical expectations (see \S \ref{sect:expansion}), because disk expansion is set by the physics of the outer, externally-irradiated disk regions, in which both $T$ and $\nu$ are independent of $M_d$, see equation (\ref{eq:T_irr}).

%%%%%%%%%%%%%%%%%%%%%%%%%%%%%%%%%%%%%%%%%%%%%%%%%%%%%%%%%%%

\subsection{Evolution of $\FJ$ distribution.}  
\label{sect:FJ}

\begin{figure}
\plotone{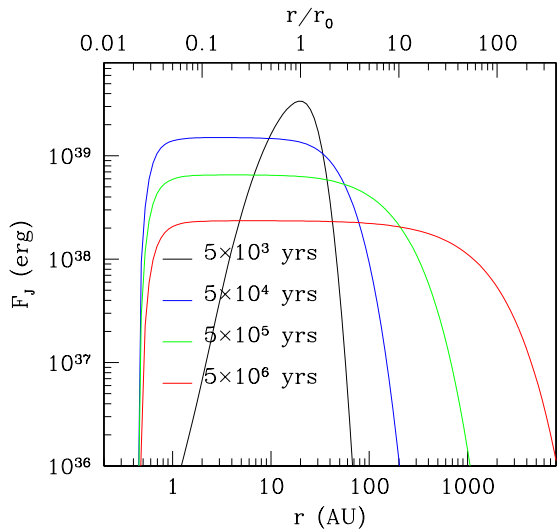}
\caption{Evolution of $\FJ(r,t)$ in a circumbinary disk with $M_d=0.1 M_c$. Different curves correspond to different epochs: black --- $5.0 \times 10^{3}$ yrs, blue --- $5.0 \times 10^{4}$ yrs, green --- $5.0\times 10^{5}$ yrs, red --- $5.0 \times 10^{6}$ yrs. At time $t=0$ mass is distributed in a narrow Gaussian ring at radius $r_0=20$ AU, see equation (\ref{eq:initial-cond}). It subsequently spreads viscously both inward and outward of $r_0$. Note the overall expansion of the disk with time and formation of the central cavity. }
\label{fig:FJ}
\end{figure}

We now explore the evolution of $\FJ$ distribution in our numerical models. Figure \ref{fig:FJ} shows $\FJ(r,t)$ at different moments of time for a disk with $M_d=0.1 M_c$. 

Initially the disk evolves on a timescale considerably shorter (by an order of magnitude) than the viscous time $t_\nu=r_0^2/\nu$ at $r_0$, which is in the irradiated part of the disk. Figure \ref{fig:FJ} shows that already at $5\times 10^3$ yr $\approx 0.01t_\nu(r_0)$ (black curve) viscous mass inflow reaches the central binary, where it is stopped by the tidal torque, giving rise to a plateau in the radial distribution of $\FJ$ at $r$ of order several AU. Such rapid evolution is an artefact of the strong radial gradients of $\Sigma$ associated with its initial narrowly-peaked distribution in the form of a ring, see equation (\ref{eq:initial-cond}). Sharp drop of $\FJ$ at small radii is caused by the disk truncation by the binary torques and formation of the central cavity. At larger separations $r\sim r_0$ relaxation to $\FJ=\,$const state has not yet been achieved, and mass flows both inward and outward from the $r\sim r_0$ region. 

%%%%%%%%%%%%%%%%%%%
%%%%%%%%%%%%%%%%%%%

At $t=7\times 10^4$ yr $\approx 0.16t_\nu(r_0)$ (blue curve) radial 
distribution of $\FJ$ becomes consistent with being constant over more 
than an order of magnitude in radius. This implies that $\dot M$ in
this part of the disk is very small. At later times viscous 
expansion of the disk extends the outer disk edge by about an order
of magnitude, generally preserving radially constant $\FJ$ over the large span of the 
disk.

\begin{figure}
\plotone{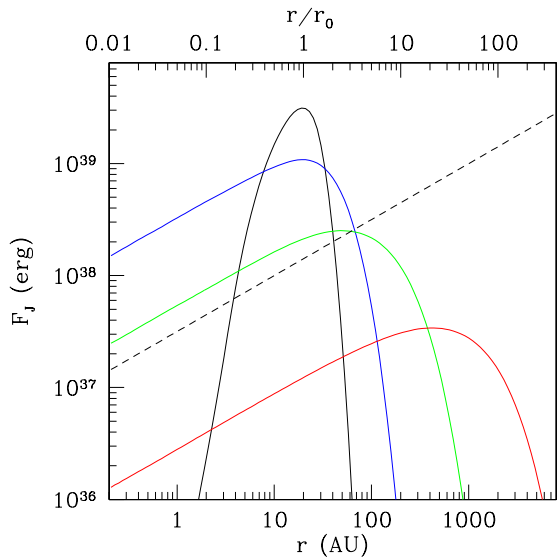}
\caption{Same as Figure \ref{fig:FJ} but for a circumstellar disk of $M_d=0.1 M_c$ around a single star with the same total central mass as in circumbinary case. Different colors correspond to the same moments of time as in Figure \ref{fig:FJ}:  black --- $5.0 \times 10^{3}$ yrs, blue --- $5.0 \times 10^{4}$ yrs, green --- $5.0\times 10^{5}$ yrs, red --- $5.0 \times 10^{6}$ yrs. However, the structure of the radial distribution of $\FJ$ is very different compared to the case of a circumbinary disk.}
\label{fig:FJ-circumstellar}
\end{figure}

At the same time, the amplitude of $\FJ$ in the bulk of the disk (the height of the plateau) constantly decreases at late time, in agreement with theoretical expectations (\S \ref{sect:F_J_evol}). To quantify this behavior we measure $F_J^{\rm pl}$ --- the height of the plateau of the $\FJ(r,t)$ distribution in our simulations --- for three different disk masses at different moments of time and plot it in Figure \ref{fig:FJ-circumstellar}. It is clear that at late stages, $t\gtrsim t_\nu(r_0)$, $\FJ$ decays roughly as a power law in time. Also, more massive disks feature higher values of $\FJ$. 

By fitting these numerical results we found that at late times
\ba
F_J^{\rm pl}(t) \approx 0.45 \FJ(t),
\label{eq:FJ_via_M_d_num}
\ea
where $\FJ(t)$ is given by equation (\ref{eq:FJ_via_M_d}). This verifies and calibrates our
simple scaling (\ref{eq:FJ_via_M_d}) for the behavior of $\FJ$ presented in \S \ref{sect:F_J_evol}, and motivates the use of our analytical results in other circumbinary disk applications.

%%%%%%%%%%%%%%%%%%%%%%%%
\begin{figure}
\plotone{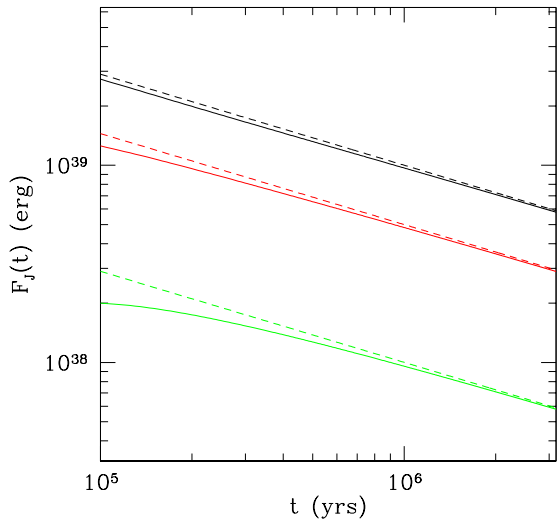}
\caption{Time evolution of $F_J^{\rm pl}(t)$ --- value of $\FJ$  at the plateau portion of the $\FJ(r)$ distribution in our simulations of circumbinary disks around non-accreting central binary. Different colors correspond to different disk masses: $M_d=0.1 M_c$ (black), $0.05 M_c$ (red), and $0.01 M_c$ (green). Solid curves are the results of our numerical calculations, while dashed lines represent our analytical fit given by equations (\ref{eq:FJ_via_M_d}) \& (\ref{eq:FJ_via_M_d_num}).}
\label{fig:ftime}
\end{figure}
%%%%%%%%%%%%%%%%%%%%%%%%

Evolution of $\FJ$ distribution shown in Figure \ref{fig:FJ} should be contrasted with $\FJ(r,t)$ behavior in the case of a circumstellar disk around a single star without angular momentum injection at the center. Figure \ref{fig:FJ-circumstellar} displays how $\FJ$ evolves in the latter case, keeping the same central mass $M_c$ and all disk parameters. Profiles of $\FJ$ are plotted at the same moments of time as in Figure \ref{fig:FJ}. One can see a dramatic difference in $\FJ$ behavior compared to the circumbinary case: in circumstellar disk $\FJ(r)$ never becomes flat. Instead, $\FJ$ distribution rapidly converges to a standard constant $\dot M$ solution (\ref{eq:const_Mdot}) with $F_{J,0}=0$, i.e. $\FJ(r)\propto r^{1/2}$.

The amplitude of $\FJ$ in Figure \ref{fig:FJ-circumstellar} rapidly drops in time for two reasons. First, analogous to the circumbinary case, the disk expands, lowering $\Sigma$ and, consequently, reducing $\FJ$ (this effect operates in circumbinary disks as well). Second, ongoing mass accretion onto the central object (absent in the circumbinary disk with $\dot M_b=0$) reduces disk mass on global viscous timescale and additionally lowers $\FJ$. Lower value of $\FJ$ in a circumstellar disk, especially in its inner regions, diminishes the role of viscous heating in determining its thermal state compared to the case of a circumbinary disk.

%%%%%%%%%%%%%%%%%%%

%%%%%%%%%%%%%%%%%%%%%%%%%%%%%%%%%%%%%%%%%%%%%%%%%%%%%%%%%%%

\subsection{Evolution of disk properties.}  
\label{sect:disk_prop}

\begin{figure}
\epsscale{1.1}
\plotone{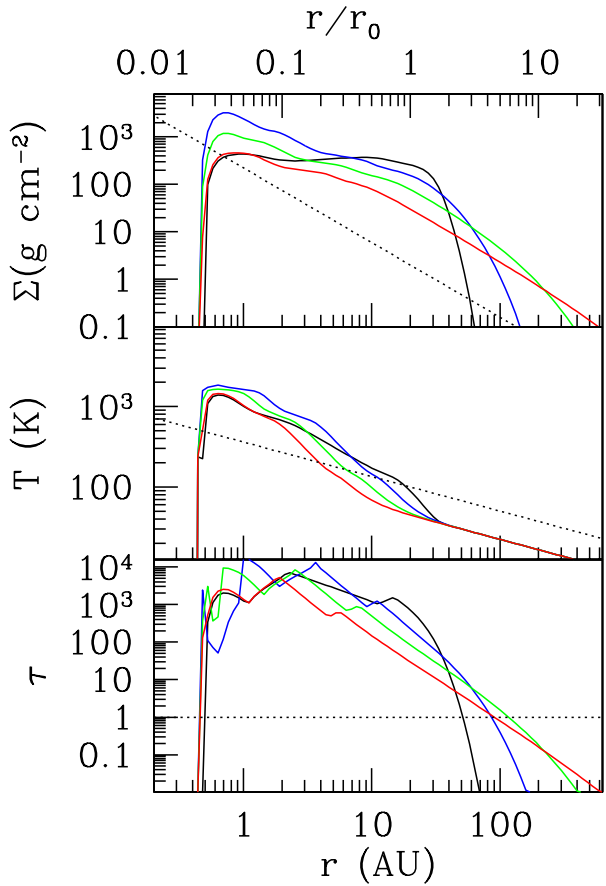}
\caption{Evolution of the surface density (top), midplane temperature (middle), and optical depth (bottom) in our numerical models of a high-mass circumbinary disk with $M_d=0.1M_c$ without mass accretion at the center, $\dot M_b\approx 0$. System with the following parameters is shown: equal mass ($q=1.0$) central binary of total mass $M_{c}=M_{\odot}$ and luminosity $L_{c} = 2L_{\odot}$. For this disk mass we set $f=2 \times 10^{-3}$ in the torque density prescription (\ref{eq:Lambda}) for the central cavity to have a radius $r_c\approx 2a_b$, in agreement with hydro simulations. Different colors correspond to different times: black $5.0 \times 10^{3}$ yrs, blue $5.0 \times 10^{4}$ yrs, green $5.0 \times 10^{5}$ yrs, red $5.0 \times 10^{6}$ yrs. The dotted lines in two upper panels correspond to the analytic scalings for $\Sigma(r)$ and $T(r)$ for the irradiation-dominated case, equations (\ref{eq:T_irr}) and (\ref{eq:surf_irr}). Note the relatively slow evolution of $\Sigma(r)$ caused by the assumed lack of accretion onto the binary.}
\label{fig:md0.1}
\end{figure}

\begin{figure}
\epsscale{1.1}
\plotone{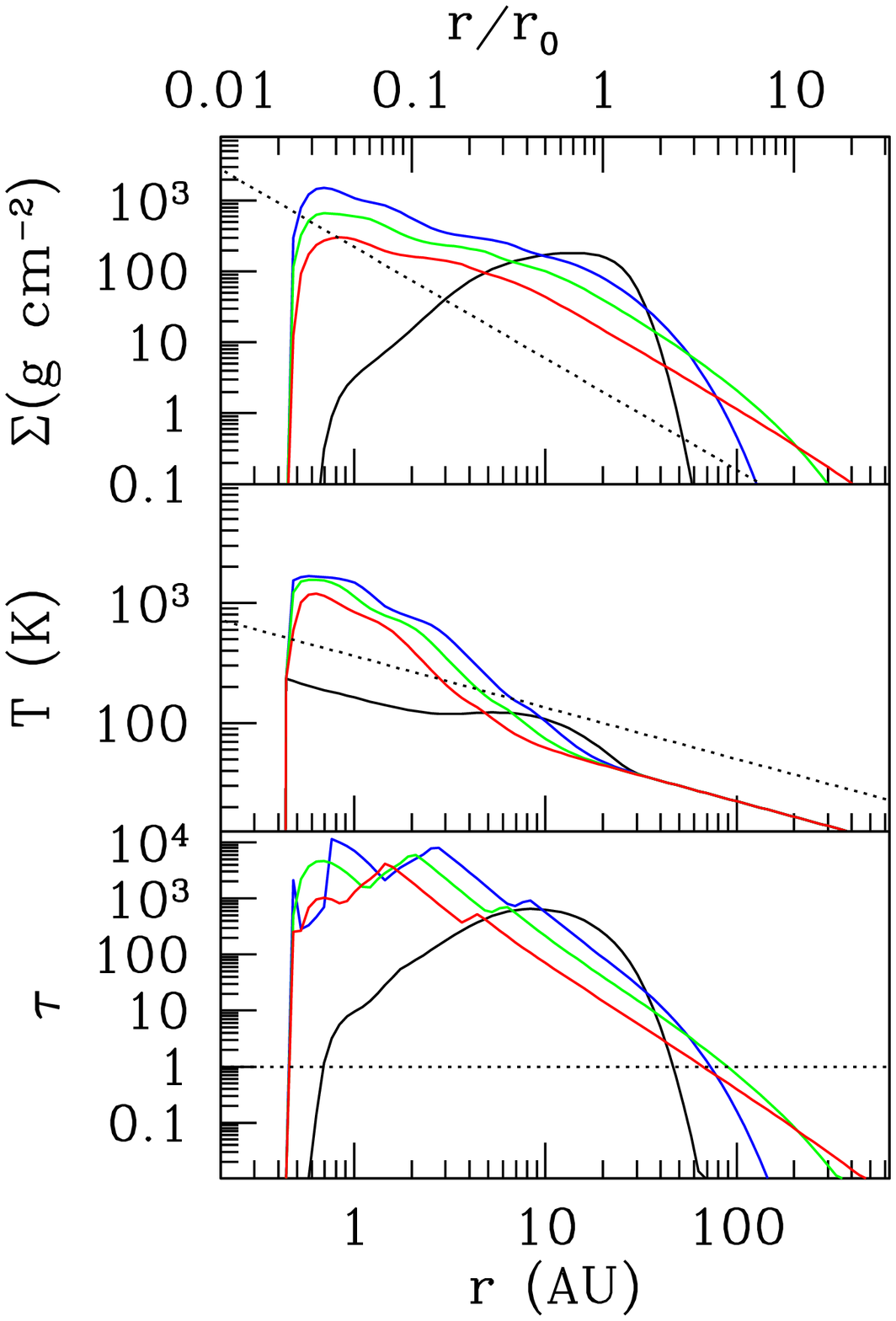}
\caption{Same as Figure \ref{fig:md0.1} but for $M_d=0.05M_c$ and $f=1.5 \times 10^{-3}$.}
\label{fig:md0.05}
\end{figure}

\begin{figure}
\epsscale{1.1}
\plotone{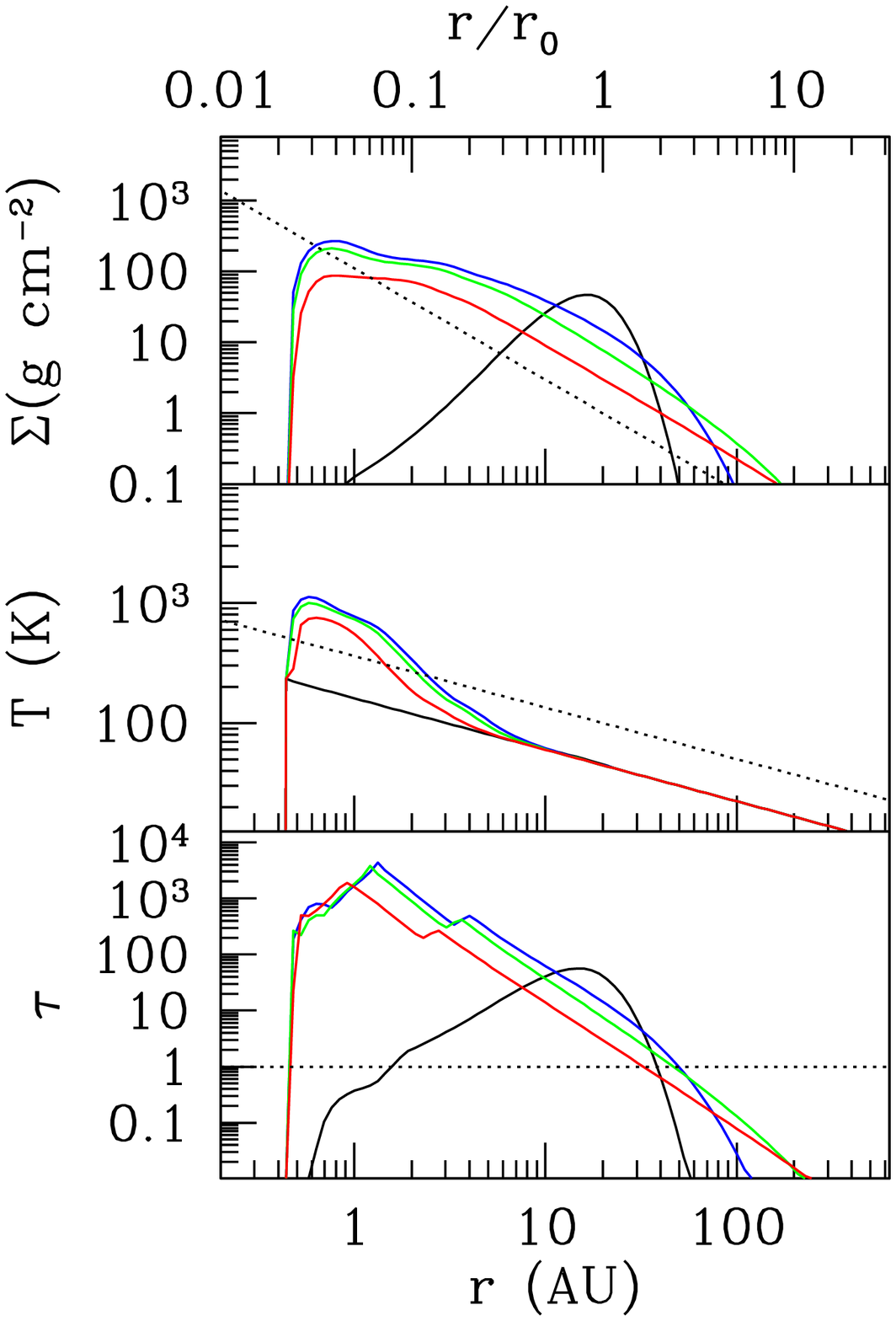}
\caption{Same as Figure \ref{fig:md0.1} but for $M_d=0.01M_c$ and $f=1.0 \times 10^{-3}$.}
\label{fig:md0.01}
\end{figure}

We will now look at how other basic disk properties evolve in our numerical models. Figures \ref{fig:md0.1}-\ref{fig:md0.01} summarize our results for
three different disk masses, $M_d=0.1M_c$, $0.05M_c$, and $0.01M_c$, correspondingly. Each figure shows the surface density $\Sigma(r)$, midplane temperature $T(r)$,
and optical depth $\tau(r)$ at the same moments of time as in Figure \ref{fig:FJ}. 

Examination of these figures reveals several features common to circumbinary disks of different masses. The overall shapes of $\Sigma(r)$ curves are similar in different plots, despite the overall scales being different at the same moments of time: higher $\Sigma$ for more massive disks, and vice versa. Surface density increases towards smaller radii and exhibits a clear signature of a central cavity that is placed at $\sim 2a_b$ by design (by choosing a proper value of the coefficient $f$ in our torque density prescription (\ref{eq:Lambda})), indicated in each Figure. Peak values of $\Sigma(r)$ at each moment of time scale roughly linearly with $M_d$ used in the calculation.

At the same time, temperature profiles for different disk masses vary less significantly, with the difference in peak values of $\sim 2$ between the high- and low-mass disk models ($M_d=0.1M_c$ and $0.01M_c$, respectively). Moreover, beyond $10-30$ AU the midplane temperature converges to the prediction (\ref{eq:T_irr}) at all times, because central irradiation becomes the dominant factor setting $T$ (see \S \ref{sect:heating_terms}). For the same reason $\Sigma(r)$ on these scales obeys equation (\ref{eq:surf_irr}) at late times (dotted lines), see the red curve at 5 Myrs in top panels of Figures \ref{fig:md0.1}-\ref{fig:md0.01}.

Viscous dissipation plays important role in setting midplane temperature of the inner disk regions. It dominates over irradiation for $r\lesssim 10-30$ AU for $M_d=(0.05-0.1)M_c$, while for $M_d=0.01M_c$ its effect is confined to within several AU from the binary. High-mass disks can be warmed up by viscous and tidal dissipation to $2\times 10^3$ K in their inner regions, while in low mass disks $T$ does not exceed $10^3$ K.

Optical depth of the disk (lower panels of Figures \ref{fig:md0.1}-\ref{fig:md0.01}) exhibits complicated structure related to sublimation of different grain species responsible for the opacity of the disk \citep{zhu_2009}. These opacity transitions are one of the reasons for the fine structure visible in $\Sigma(r)$ and $T(r)$ profiles. They become less numerous at lower disk masses since viscous heating and tidal dissipation do not raise midplane temperature to very high values necessary for sublimating more refractory grains in this case.

All our numerical models become optically thin beyond $40$-$100$ AU, depending on $M_d$. There $T(r)$ may differ slightly from the prediction (\ref{eq:T_irr}), see the discussion after equation (\ref{eq:T_thin}). However, the correction --- a weak power of the ratio of grain emissivity at the peak wavelengths of the blackbody emission of the superheated layer and the disk midplane \citep{chiang_1997}, set to be the same in our simple approximation --- is not very different from unity and varies with $r$ very weakly to change our results significantly.

\begin{figure}
\epsscale{1.1}
\plotone{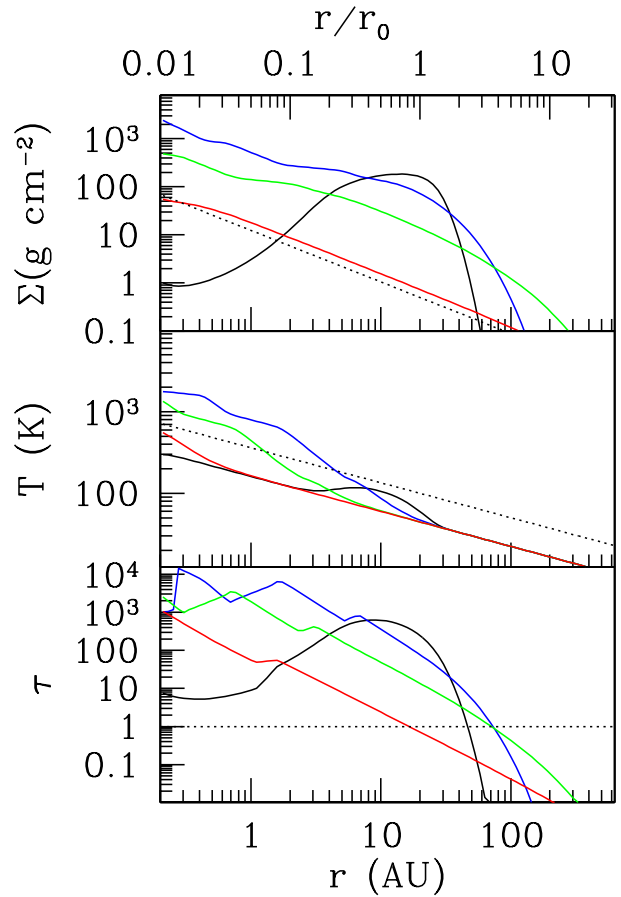}
\caption{Same as Figure \ref{fig:md0.05} but for a circumstellar disk with non-zero accretion onto the star but no external angular momentum injection. Dotted line in the upper panel illustrates $\Sigma(r)\propto r^{-15/14}$ behavior typical for irradiated disks. See \S \ref{sect:disk_prop} for details.}
\label{fig:md0.05_circumstellar}
\end{figure}

Now we provide comparison with the case of a circumstellar disk without external sources of angular momentum. In Figure \ref{fig:md0.05_circumstellar} we show the evolution of $M_d=0.05M_c$ {\it circumstellar} disk; it should be compared with Figure \ref{fig:md0.05} which illustrates the evolution of a circumbinary disk with $\dot M_b\approx 0$ of the same mass. In computing the evolution of a circumstellar disk we simply set $\Lambda=0$, keeping everything else the same as in the calculation of Figure \ref{fig:md0.05}. 

Lack of central cavity is not the only obvious feature of the circumstellar disk. Comparing Figures \ref{fig:md0.05_circumstellar} and \ref{fig:md0.05} we see that $\Sigma$ drops much faster with time in a circumstellar disk. Also, the slope of $\Sigma(r)$ profile is shallower in the circumstellar case, see the red curve in top panel. It obeys $\Sigma\propto r^{-15/14}$, which can be obtained from equation (\ref{eq:surf_irr}) by substituting $\FJ\propto r^{1/2}$ as appropriate for a circumstellar disk. These differences are easily understood in terms of the very different evolution of the $\FJ(r,t)$ distributions shown in Figures \ref{fig:FJ} and \ref{fig:FJ-circumstellar}: in the circumstellar case $\FJ$ (and, subsequently, $\Sigma$) is much lower in the inner disk and drops with time more rapidly due to ongoing accretion onto the central object. Because of that at late time ($t\sim$ 5 Myr) $T(r)$ in circumstellar disk is set entirely by irradiation outside of $\sim 0.5$ AU (compared to $\sim 10$ AU in a circumbinary case).

Based on that we conclude that circumbinary disks are very different from their circumstellar analogues: they tend to remain more massive and maintain high levels of viscous dissipation, which keeps them hotter for longer.

%%%%%%%%%%%%%%%%%%%%%%%%%%%%%%%%%%%%%%%%%%%%%%%%%%%%%%%%%%%%%%%%

\section{Spectra of circumbinary disks.}
\label{sect:spectra}

%%%%%%%%%%%%%%%%%%%%%%%%%%%%%%%%%%%%%%%%%%%%%%%%%%%%%%%%%%%%%%%%

Given significant differences in properties of circumbinary and circumstellar protoplanetary disks outlined in previous section one should expect their spectral energy distributions (SEDs) to also differ. We describe the details of our SED calculations in Appendix \ref{ap:SED} and provide comparison between SEDs of two types of disks in this section.

\begin{figure}
\epsscale{1.1}
\plotone{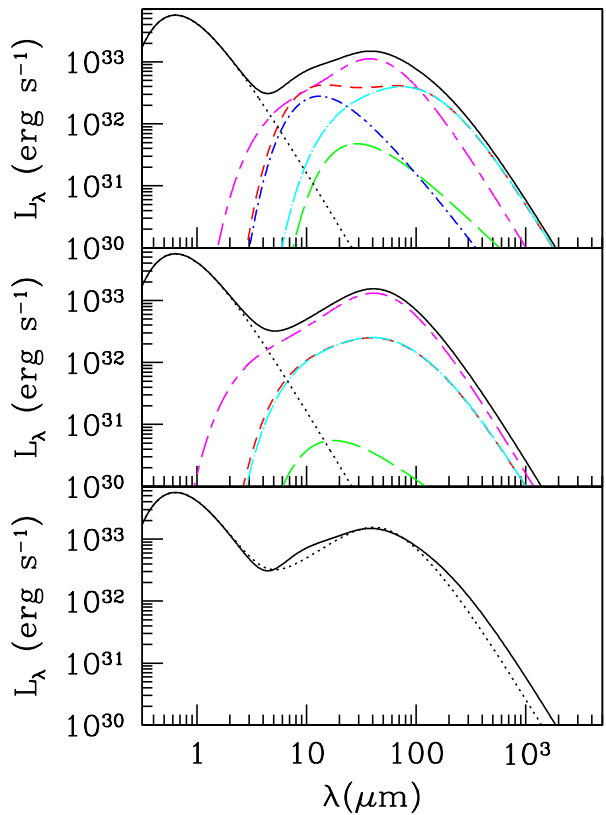}
\caption{SEDs of a circumbinary (top panel) and circumstellar (middle panel) disk with $M_d=0.05 M_c$. Different colors correspond to 
different contributions to the SED: solid black (total), dotted black (stellar), violet (superheated grains), and
red (disk). The disk contribution is further divided into its components: blue (tidal heating), cyan (irradiation), and green (viscous heating). The middle panel, corresponding to the SED of a circumstellar disk, has no tidal component. The bottom panel compares the total circumbinary SED (solid black line) with the total circumstellar SED (dotted black line).}
\label{fig:SED}
\end{figure}

Figure (\ref{fig:SED}) shows the SEDs of the circumbinary (top panel) and circumstellar (middle panel) disks with starting mass $M_d=0.05 M_c$ at $t=5$ Myr. In these plots we separately show the contributions from the stellar blackbody (dotted), superheated dust layer (long-short dashed), and the bulk of the disk (dashed). The last component is also split into tidal (only for circumbinary case), viscous, and irradiation contributions. Each of these is evaluated by setting the other heating sources to zero\footnote{I.e. tidal contribution is obtained by setting ${\cal F}_v={\cal F}_{\rm irr}=0$ in equations (\ref{eq:T_e}) and (\ref{eq:SED}); other contributions are obtained similarly.}. 

The bottom panel of Figure (\ref{fig:SED}) provides a direct comparison of the total SEDs obtained for the two disks. Its inspection highlights three important differences. First, the circumstellar disk is a stronger emitter at $\lambda\sim 3\mu$m simply because it gets closer to the star and reaches higher temperatures. Presence of the central cavity suppresses the near-IR emission of the circumbinary disk. 

Second, circumbinary SED exhibits a bump at $\lambda\sim 10\mu$m, which is absent in circumstellar SED. This feature arises because of the tidal dissipation in a circumbinary disk, which dominates heating of its inner parts (see \S \ref{sect:heating_terms}). Note that in our SED calculation the significance of the bump is reduced by the high level of the superheated dust emission, which dominates because we neglect the fact that upper layers of the disk should become optically thin to stellar radiation at large separations. If we accounted for this effect, the superheated dust contribution would be lower, giving rise to larger difference between the circumbinary and circumstellar SEDs around $10\mu$m. 

Third, circumbinary disk provides more emission in the Rayleigh-Jeans tail. This is because at $t=5$ Myr this disk has more mass than its circumstellar counterpart, which has already lost $\approx 82 \%$ of its original mass to accretion onto the central star by that time.

The distinctive bump in the SED at $\lambda\sim 10\mu$m may serve as a circumbinary disk signature that could be used to identify such systems in observations, when the binarity of the central source is difficult to determine. Such SED-based technique using features in quasar spectra has been previously suggested as a way of finding binary supermassive black holes in centers of galaxies \citep{rafikov_2013c,Yan}. In practice, in protostellar case such identification may be complicated by the presence of a strong silicate dust resonance in the disk emission at $9.7\mu$m \citep{mathis,decolle} (we do not account for it here), which could lead to ambiguity in interpretation.

%%%%%%%%%%%%%%%%%%%%%%%%%%%%%%%%%%%%%%%%%%%%%%%%%%%%%%%%%%%%%%%%

\subsection{Disk shadowing}
\label{sect:shadowing}

\begin{figure}
\plotone{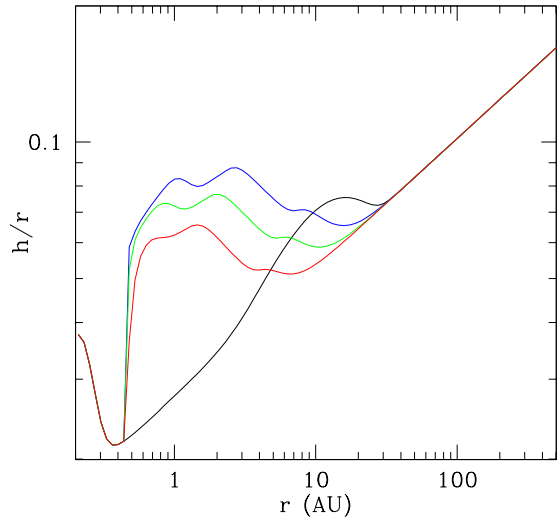}
\caption{Aspect ratio of $0.05 M_{\odot}$ circumbinary disk. Different
      colors correspond to different times: black $5.0 \times 10^{3}$ yrs, blue
      $5.0 \times 10^{4}$ yrs, green $5.0
      \times 10^{5}$ yrs, red $5.0 \times 10^{6}$ yrs. The disk is self-shadowed at
	  1-20 AU.}
\label{fig:hr}
\end{figure}

Our detailed calculations of the thermal structure of circumbinary disks in \S \ref{sect:disk_prop} reveal an interesting effect, which is illustrated in Figure \ref{fig:hr}. In this Figure we plot the disk aspect ratio $h/r$ as a function of radius at different moments of time. The outer parts of the disk are dominated by central irradiation and $h/r$ behavior there is well described by equation (\ref{eq:hr_irr}), resulting in a {\it flared} disk structure. However, at $r\lesssim 10$ AU situation changes and $h/r$ stays roughly constant or even increases with decreasing $r$. In our calculations this pattern persists at all times (except the earliest epochs affected by the initial conditions) and implies that between 1 and 10 AU circumbinary disk should have {\it non-flared} structure. 

This behavior is caused by the vigorous tidal and viscous heating in the inner disk, as we demonstrate in detail in \S \ref{sect:heating_terms}. This energy release dramatically increases midplane temperature of the disk and boosts $h/r$. This feature is unique for circumbinary disks (i.e. it is not present in the circumstellar disks) because of the tidal heating and peculiar $\FJ(r)$ behavior endemic for such disks. 

Deviation of $h/r$ behavior from equation (\ref{eq:hr_irr}) means that our procedure for calculating ${\cal F}_{\rm irr}$ (which assumes $\zeta$ to be given by equation (\ref{eq:zeta_irr}), see \S \ref{sect:setup}) becomes inaccurate at $r\lesssim 10$ AU. Moreover, in regions where $h/r$ is lower than in some interior part of the disk, that inner portion of the disk blocks stellar irradiation from reaching the disk surface and ${\cal F}_{\rm irr}\to 0$, resulting in {\it self-shadowing} of the disk. Accounting for this effect is greatly complicated by its global nature of the effect. Accurate description of the self-shadowing may necessitate direct radiation transfer calculations.

Both our evolutionary and SED calculations presently include reprocessed stellar radiation emitted from disk regions that are in fact shadowed. This is not a serious issue in the inner disk ($\lesssim 10$ AU) simply because there ${\cal F}_{\rm irr}$ is subdominant relative to both ${\cal F}_{v}$ and ${\cal F}_{\rm tid}$. Thus, retaining irradiation contribution in the energy budget produces negligible effect in this region.

More worrisome could be the fact that the inner disk regions may cast shadow over significant portion of the disk at intermediate separations where we currently consider central irradiation to dominate over the viscous and tidal heating. For example, focusing on $0.5$ Myr curve in Figure \ref{fig:hr} we see that 
aspect ratio peaks at the level of $h/r\approx 0.07$ at $r\approx 2$ AU. Even though ${\cal F}_{\rm irr}$ starts to dominate $h/r$ behavior around 10 AU, aspect ratio again reaches 0.07 in the irradiation-dominated outer disk only around $20-30$ AU. This implies that the geometrically-thick parts of the disk at 2 AU cast shadow out to $20-30$ AU at this moment of time. This is likely to lower temperature in this part of the disk and reduce irradiation contribution\footnote{This would additionally accentuate $10\mu$m bump in the circumbinary SED discussed in \S \ref{sect:spectra}.} to the SED coming from that region. While we neglect self-shadowing in this work, future studies should account for its effect on the disk structure and observational appearance.

%%%%%%%%%%%%%%%%%%%%%%%%%%%%%%%%%%%%%%%%%%%%%%%%%%%%%%%%%%%
%%%%%%%%%%%%%%%%%%%%%%%%%%%%%%%%%%%%%%%%%%%%%%%%%%%%%%%%%%%

\section{Discussion.}  
\label{sect:disc}

%%%%%%%%%%%%%%%%%%%%%%%%%%%%%%%%%%%%%%%%%%%%%%%%%%%%%%%%%%%

We now discuss some additional aspect of the circumbinary disk evolution which follow from the results presented in \S \ref{sect:numerical}.

%%%%%%%%%%%%%%%%%%%%%%%%%%%%%%%%%%%%%%%%%%%%%%%%%%%%%%%%%%%

\subsection{Role of different heating terms.}  
\label{sect:heating_terms}

%%%%%%%%%%%%%%%%%%%%%%%%%%%%%%%%%%%%%%%%%%%%%%%%%%%%%%%%%%%

\begin{figure}
\plotone{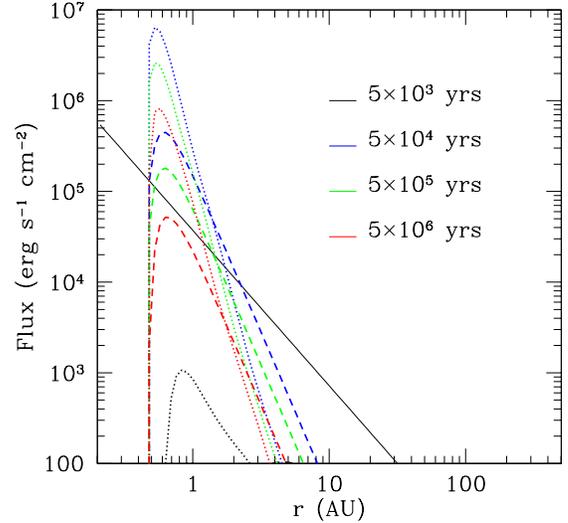}
\caption{Relative contributions of different heating terms for a $0.05 M_{\odot}$ circumbinary disk: viscous heating (dashed), tidal dissipation (dotted), and irradiation (solid black). From top to bottom the times are: $5.0 \times 10^3$ yr (black),
$5.0\times 10^4$ yr (blue), $5.0 \times 10^5$ yr (green), $5.0 \times 10^6$ yr (red). Irradiation does not evolve with time and, at late times, viscous heating is dwarfed by irradiation for all radii. Tidal heating dominates in the inner disk.}
\label{fig:tvi}
\end{figure}

Here we explore the role of different heating sources in determining the thermal state of a circumbinary disk at different radii. Figure \ref{fig:tvi} shows the runs of viscous ${\cal F}_v(r)$, tidal ${\cal F}_{\rm tid}(r)$, and irradiation ${\cal F}_{\rm irr}(r)$ heating at different moments of time for $M_d=0.05M_c$ disk. Several important conclusions can be drawn from this plot. 

First, on scales $r\lesssim 2$ AU dissipation of the density waves launched by the binary provides the most important heating source. At its peak (around 0.6 AU) ${\cal F}_{\rm tid}$ exceeds ${\cal F}_v$ by about an order of magnitude at all times. To understand this behavior we will use the fact that viscous time in the inner disk is much shorter than the global evolution timescale, so that a quasi-steady state must develop there. In this limit the binary torque near the disk edge is balanced by the gradient of the viscous angular momentum flux, so that $\partial \FJ/\partial r=2\pi r\Sigma\Lambda$. We can then use this relation and equations (\ref{eq:flux_nu}) and (\ref{eq:ftid}) to write
\ba
\frac{{\cal F}_{\rm tid}}{{\cal F}_v}=\frac{2}{3}\frac{\partial\ln\FJ}{\partial\ln r}\frac{\Omega_b-\Omega}{\Omega}\approx \frac{2}{3}\frac{\partial\ln\FJ}{\partial\ln r}\left(\frac{r}{a_b}\right)^{3/2},
\label{eq:Frat}
\ea
where the last approximation follows from the fact that $\Omega\lesssim\Omega_b$ everywhere in the disk. 

Looking at the Figure \ref{fig:FJ} we see that $\partial\ln\FJ/\partial\ln r$ is very large near the inner edge of the disk. This, coupled with the fact that $(r/a_b)^{3/2}\gg 1$, explains the major role of ${\cal F}_{\rm tid}$ in heating the disk within $\sim 2$ AU. Outside this region $\partial\ln\FJ/\partial\ln r$ rapidly decreases ($\FJ$ distribution flattens, see Figure \ref{fig:FJ}), and viscous dissipation starts to dominate over the tidal heating.

The dominant role of the tidal heating in the inner part of the disk makes it important to understand the detailed radial structure of the density wave dissipation in the disk, since ${\cal F}_{\rm tid}$ is directly connected to $\Lambda(r)$. Our simple prescription (\ref{eq:Lambda}) ignores both the recent developments in understanding the excitation torque density \citep{RP12,petrovich_2012} and the non-local nature of the density wave damping \citep{GR01,R02}, motivating further refinements. 

\begin{figure}
\plotone{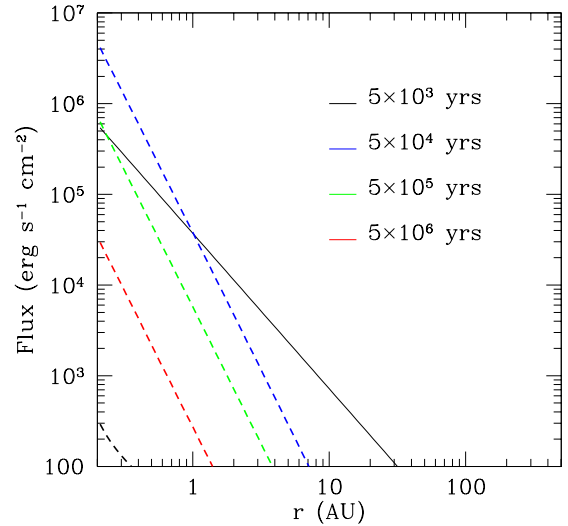}
\caption{Same as Figure \ref{fig:tvi} but for a $0.05 M_{\odot}$ circumstellar disk (note the lack of tidal heating). Viscous heating is less significant in the circumstellar case than in the circumbinary case because of the pile-up of mass at small radii in the latter case.}
\label{fig:tvics}
\end{figure}

Second, irradiation by the binary is always the major heating source at large separations. This is because ${\cal F}_v\propto r^{-7/2}$ in a constant-$\FJ$ disk, while ${\cal F}_{\rm irr}\propto \zeta(r)/r^2$ decays much slower with $r$. Because both viscous heating and tidal dissipation scale with $M_d$, we expect irradiation to dominate closer to the star for less massive disks. 

Third, even when ${\cal F}_{\rm irr}\gtrsim {\cal F}_v,{\cal F}_{\rm tid}$ tidal and viscous dissipation may not be neglected when calculating the midplane disk temperature $T$. For example, Figure \ref{fig:tvi} shows that ${\cal F}_{\rm irr}\approx  {\cal F}_v+{\cal F}_{\rm tid}$ around 1.5 AU at $t=5$ Myr. However, Figure \ref{fig:md0.05} demonstrates that $T$ converges to the behavior (\ref{eq:T_irr}) only outside $8$ AU. This is because ${\cal F}_v$ and ${\cal F}_{\rm tid}$ have a weighting factor $\tau$ in equation (\ref{eq:gen_T}) determining $T$, and the disk is optically thick inside 10 AU.

In Figure \ref{fig:tvics} we plot ${\cal F}_{\rm irr}$ and ${\cal F}_v$ in a circumstellar disk of the same mass as in Figure \ref{fig:tvi}. Comparing the two Figures one can see that in the circumstellar case ${\cal F}_v$ becomes a subdominant heating source faster than in the circumbinary case, as a result of accretion onto the central star in the former case. Also, the ${\cal F}_v(r)$ profile is shallower in the circumstellar case, because $\FJ\propto r^{1/2}$ for a constant-$\dot M$ disk (see \S \ref{sect:FJ}) and equation (\ref{eq:flux_nu}) then predicts ${\cal F}_v(r)\propto r^{-3}$, as opposed to ${\cal F}_v\propto r^{-7/2}$ in a constant-$\FJ$ circumbinary disk.

%%%%%%%%%%%%%%%%%%%%%%%%%%%%%%%%%%%%%%%%%%%%%%%%%%%%%%%%%%%%%%%%

\subsection{Effect of accretion onto the binary}
\label{sect:someaccretion}

In most of this work we assumed that the binary torque cuts off disk inflow into the central cavity and $\dot M_b\approx 0$. At the same time, \cite{artymowicz_1996} suggested that tidal streams could penetrate the cavity, giving rise to some accretion by the binary.  Early work \citep{mcfadyen_2008} suggested that $\dot M_b$ due to tidal streams is $\sim 10\%$ the accretion rate without central torque. However, recent hydrodynamic \citep{farris_2014, orazio_2013} and MHD \citep{shi_2012} simulations suggest that $\dot M_b$ may reach $(30 - 60)\%$ of the corresponding accretion rate without central torque.

For brevity we do not show numerical disk models with $\dot M_b\neq 0$ (except or the standard constant-$\dot M$ circumstellar case) in this work. However, we outline qualitatively how our results should change in this case.

As long as the outer disk edge remains in the irradiation-dominated part of the disk equations (\ref{eq:t_nu_0}) and (\ref{eq:rinfl}) remain valid. But equation (\ref{eq:M_d}) gets modified for two reasons. First, because of accretion at some (generally time-dependent) rate $\dot M_b(t)$ disk mass is now a {\it decreasing function of time} and $dM_d/dt=-\dot M_b$. Second, $\FJ(r)$ is no longer constant with $r$ and instead obeys the solution (\ref{eq:lin_sol}), as long as variations of $\dot M_b$ are not faster\footnote{It has to be kept in mind that $\dot M_b(t)$ is likely to be set not by the global disk properties, but by its characteristics near the cavity edge, since this is where binary torque is posing a barrier to gas inflow.} than the global disk evolution. For that reason formally we can no longer take $\FJ$ out of the integral in equation (\ref{eq:M_d}). 

Nevertheless, it is obvious that this integral is still dominated by the outermost disk regions, where $\FJ$ attains its maximum value $\approx \FJ(r_{\rm infl})$. Thus, we can approximately evaluate the integral at its upper limit and find that, up to factors of order unity, equations (\ref{eq:FJ_irr}) and (\ref{eq:FJ_via_M_d}) remain valid as long as we (1) replace in them $\FJ$ with $\FJ(r_{\rm infl})$ and (2) consider disk mass to be a function of time, $M_d(t)$, which can be computed once the history of accretion onto the binary, $\dot M_b(t)$, is known. Once the approximate value of $\FJ(r_{\rm infl})$ is determined and current value of $\dot M_b$ is known, the global run of $\FJ(r)$ (for $r\lesssim r_{\rm infl}$) is given by  
\ba
\FJ(r)\approx \FJ(r_{\rm infl})+\dot M_b\left[l(r)-l(r_{\rm infl})\right].
\label{eq:FJ_mod}
\ea
With this relation one can then easily determine the profiles of $\Sigma(r)$, $\tau(r)$, etc. using definition (\ref{eq:Fnu}) in much the same way as we derived equation (\ref{eq:surf_irr}). 

This simple recipe provides a prescription for quantitative understanding of the effect of non-zero binary accretion on evolution of the circumbinary disk properties. Apparently, since the disk is losing mass, $\FJ$ is going to be smaller than in the constant-$\FJ$ circumbinary disk with $\dot M_b\approx 0$, resulting in lower-$\Sigma$, cooler, and less luminous disk. It is also clear that properties of the disk with $\dot M_b\neq 0$ will always be somewhere in between the circumstellar case with no torque (which provides the maximum $\dot M_b$) and the constant-$\FJ$ circumbinary disk, both of which have been covered in \S \ref{sect:numerical}.

%%%%%%%%%%%%%%%%%%%%%%%%%%%%%%%%%%%%%%%%%%%%%%%%%%%%%%%%%%%%%%%%

\subsection{Binary Inspiral due to Disk Coupling}
\label{sect:binangloss}

Throughout this work we assumed binary separation to remain fixed at $a_b=0.2$ AU. However, in practice, binary is constantly losing its angular momentum due to tidal coupling with the disk. Here we look at this process in some detail.

The amount of angular momentum that the binary possesses is
\ba
L_b &=& \frac{q}{(1+q)^2}\left(GM_c^3 a_b\right)^{1/2}
\label{eq:L_b}\\
%%%%%
&\approx & 4\times 10^{52}~\mbox{g cm}^2\mbox{s}^{-1}~\frac{q}{(1+q)^2}M_{c,1}^{3/2} \left(\frac{a_b}{0.2\mbox{AU}}\right)^{1/2}.
\nonumber
\ea
Binary torque provides a source of the angular momentum for the disk with "angular momentum power" equal to $F_{J,0}\approx \FJ(r_\Lambda)$, where $F_{J,0}$ is featured in quasi-steady solution (\ref{eq:lin_sol}) and $r_\Lambda$ is the radius, beyond which only viscous stresses matter (i.e. binary torque can be neglected). As a result $L_b$ changes at the rate $\dot L_b=-F_{J,0}$. 

In our constant-$\FJ$ disk with $\dot M_b\approx 0$ we have $F_{J,0}\approx \FJ^{\rm pl}$ --- the value of $\FJ$ at the plateau of its radial distribution. Using equations (\ref{eq:FJ_via_M_d}) and (\ref{eq:FJ_via_M_d_num}) based on our numerical results one can easily compute the change of $L_b$ accumulated over time $t$ to be
\ba
\Delta L_b(t) & \approx & 0.45\int\limits_0^t \FJ(t^\prime)dt^\prime \approx 0.8 \FJ(t)t
\label{eq:dL}\\
%%%%%%%%%%
&\approx & 9 \times 10^{51}~\mbox{g cm}^2\mbox{s}^{-1}~
M_{d,-2}\alpha_{-2}^{7/13}
\nonumber\\
%%%%%%%%%%
&\times &\left[\frac{M_{c,1}L_{c,1}\eta_3}
{\mu_2^4}\right]^{2/13}
\left(\frac{t}{3~\mbox{Myr}}\right)^{7/13}.
\label{eq:dL_num}
\ea

\begin{figure}
\plotone{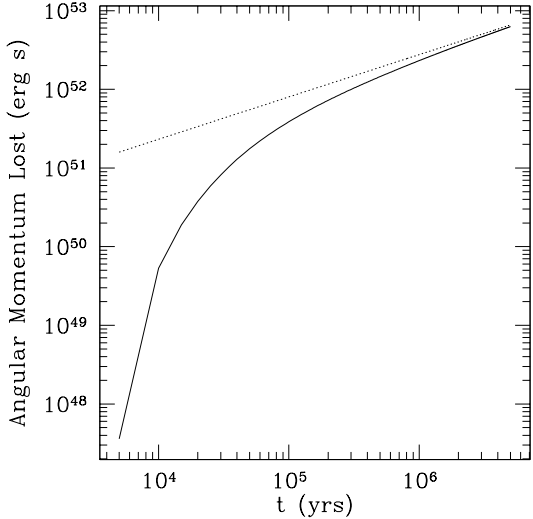}
\caption{Binary angular momentum loss due to tidal interaction with a $0.05 M_{\odot}$ circumbinary disk in the case of no central accretion, $\dot M_b\approx 0$. Dotted line shows the late-time analytical prediction for $\Delta L_b(t)$ given by equation (\ref{eq:dL_num}).}
\label{fig:angloss}
\end{figure}

In Figure \ref{fig:angloss} we show the amount of the angular momentum lost by the central binary as determined from one of our runs with $M_d=0.05M_c$ and standard binary parameters. Solid curve shows $\Delta L=2\pi\int_0^tr^\prime\Lambda(r^\prime)\Sigma(r^\prime)dr^\prime$ (angular momentum absorbed by the disk, which is equal in amplitude to the angular momentum lost by the binary) as a function of time. The dotted line is the theoretical prediction (\ref{eq:dL_num}), which fits the numerical data reasonably well at late times. The binary parameters adopted in this calculation (e.g. $q=1$) imply that initially $L_b 
\approx 10^{52}~\mbox{g cm}^2~\mbox{s}^{-1}$. Figure \ref{fig:angloss} shows that this amount of angular momentum gets lost by the binary due to tidal coupling with the disk already by $0.4$ Myr. Thus, this particular system would merge into a single star within that interval of time.

Note that our calculation of $\Delta L_b(t)$ is essentially insensitive to the binary parameters --- binary simply provides a barrier for the disk inflow causing mass accumulation at the inner edge. The global evolution of the disk is independent of the binary characteristics as long as $\dot M_b$ remains the same and $\FJ$ in the bulk of the disk stays unchanged. For that reason our analytical estimate (\ref{eq:dL_num}) does not involve binary semi-major axis, for example. Thus, we could have done the same calculation with wider or more massive binary, having higher $L_b$, and still used the curve in Figure \ref{fig:angloss} to determine whether it will merge or not (i.e. whether $\Delta L_b(t)$ ever reaches the initial $L_b$). Using this Figure we find, in particular, that all binaries with initial $L_b\gtrsim 6\times 10^{52}~\mbox{g cm}^2\mbox{s}^{-1}$ (e.g. more massive or more widely separated) would avoid merger after being embedded in a circumbinary disk for 5 Myr (although their orbits could still shrink appreciably). 

Our estimate (\ref{eq:dL_num}) has been done for the constant-$\FJ$ disk arising when $\dot M_b=0$, and it is natural to ask how it would change if some accretion onto the binary is allowed. In that case $\FJ$ profile would be represented by equation (\ref{eq:FJ_mod}), and the amount of angular momentum lost by the binary per unit of time becomes $\FJ(r\to 0)\approx \FJ(r_{\rm infl})-\dot M_b l(r_{\rm infl})$. Once the evolution of $r_{\rm infl}$ and $\FJ(r_{\rm infl})$ is specified as described in \S \ref{sect:someaccretion}, one can again compute the angular momentum lost by the binary simply as $\Delta L_b(t) \approx \int_0^t \FJ(r\to 0,t^\prime)dt^\prime$. It is obvious that a disk with non-zero $\dot M_b$ will absorb less angular momentum from the binary than its counterpart around a  non-accreting binary, simply because $\FJ(r\to 0,t)$ is always higher in the latter case. Thus, accreting binaries have a higher chance to survive against orbital inspiral caused by the tidal coupling to the circumbinary disk.

%%%%%%%%%%%%%%%%%%%%%%%%%%%%%%%%%%%%%%%%%%%%%%%%%%%%%%%%%%%%%%%%

\subsection{Dead zone}
\label{sect:deadzone}

%%%%%%%%%%%%%%%%%%%%%%%%%%%%%%%%%%%%%%%%%%%%%%%%%%%%%%%%%%%%%%%%

\begin{figure}
\plotone{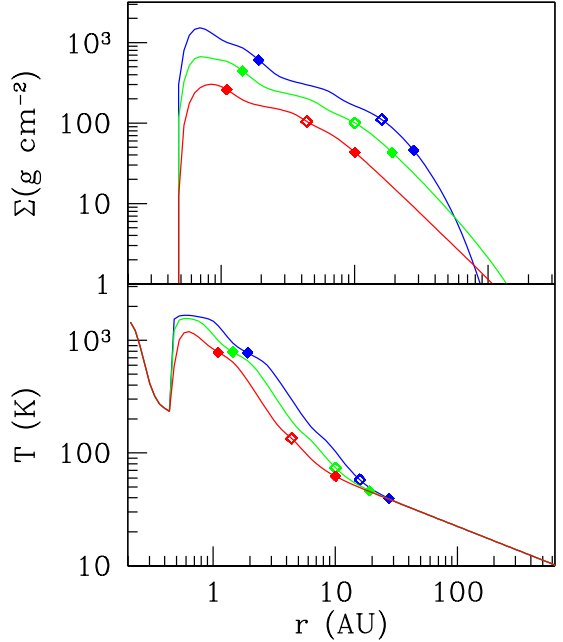}
\caption{Location of the dead zone for a circumbinary disk of mass $0.05 M_{\odot}$ at different moments of time: blue $5.0 \times 10^{4}$ yrs, green $5.0\times 10^{5}$ yrs, red $5.0 \times 10^{6}$ yrs. Filled diamonds on each $\Sigma(r)$ and $T(r)$ curve represent the inner and outer edges of the dead zone for  $\Sigma_{crit} = 40\,\mathrm{g\,cm}^{-2}$. Open diamonds illustrate the outer boundary of the dead zone for $\Sigma_{crit} = 100\,\mathrm{g\,cm}^{-2}$. Critical temperature at which thermal ionization becomes effective is taken to be 800 K in this calculation.}
\label{fig:dz}
\end{figure}

Following the work of \citet{martin_2013} we investigate the possibility of the dead zone \citep{gammie_1996} formation in our circumbinary disk models. We assume MRI to be inactive and a dead zone to form provided that the midplane temperature $T < 800$ K and disk surface density $\Sigma >\Sigma_{crit} \approx 100\,\mathrm{g\,cm^{-2}}$ \citep{gammie_1996, umebayashi_1981, zhu_2009}. At temperatures this low thermal ionization is inefficient, while the surface density constraint guarantees that cosmic ray ionization is not effective either. It has to be kept in mind that the exact value of $\Sigma_{crit}$ is not well known  \citep{sano_2008, stone_2011, zhu_2009} and we consider it as a free parameter in our study. 
We display the location and evolution of the dead zone for a $M_d=0.05 M_{\odot}$ circumbinary disk in Figure \ref{fig:dz}. At each snapshot the inner and outer boundaries of the dead zone for $\Sigma_{crit} =40$ g cm$^{-2}$ are shown as filled diamonds of the corresponding color. For higher $\Sigma_{crit} =100$ g cm$^{-2}$ the outer boundary of the dead zone shifts inward (open diamonds). We typically find that the dead zone should be present in the circumbinary disk and span several AU in radius. This is consistent with the findings of \citet{martin_2013} even though they used considerably lower values of $\Sigma_{crit}$ than we do here. 

Because of the uncertainties related to the details of the dead zone structure ($\Sigma_{crit}$, value of the background viscosity inside the zone, etc.) our present work neglects this possibility and focuses on disk evolution with a uniform value of $\alpha$. But our results shown in Figure \ref{fig:dz} do call for the development of more accurate models of the circumbinary disks that would account for the possible existence of the dead zone.

%%%%%%%%%%%%%%%%%%%%%%%%%%%%%%%%%%%%%%%%%%%%%%%%%%%%%%%%%%%%%%%%

\subsection{Iceline}
\label{sect:iceline}

%%%%%%%%%%%%%%%%%%%%%%%%%%%%%%%%%%%%%%%%%%%%%%%%%%%%%%%%%%%%%%%%

Another important location in the disk is the {\it iceline} --- radius $r_{\rm ice}$ at which water vapor (and other volatile species) condense into solids. The position of iceline in the circumbinary disks has been previously addressed in \citet{martin_2013}, \citet{clanton_2013}, \citet{Shadmehri}, and we provide comparison with these studies in \S \ref{sect:others}. Here we mainly focus on the differences in the iceline location in the circumbinary and circumstellar disks. 

\begin{figure}
\plotone{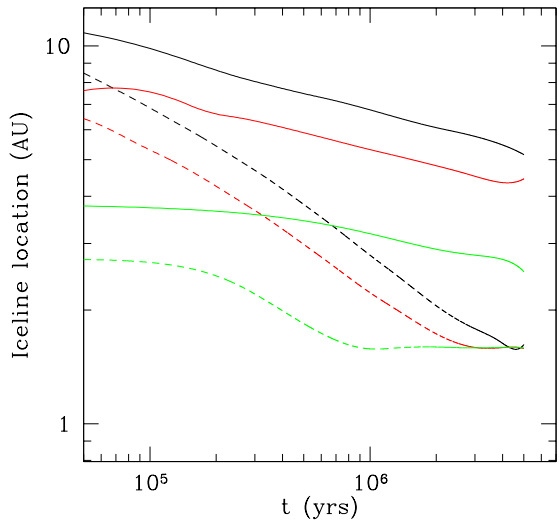}
\caption{Evolution of the iceline position $r_{\rm ice}$ for circumbinary (solid curves) and
circumstellar (dashed curves) disks. Different colors correspond to different starting disk masses: black $M_d=0.1 M_c$, red $M_d=0.05 M_c$, and green $M_d=0.01 M_c$.}
\label{fig:iceline}
\end{figure}

In Figure \ref{fig:iceline} we have plotted $r_{\rm ice}$ as a function of time for the three values of $M_d$ for the circumbinary (solid) and circumstellar (dashed) disks. Unlike most of the prior works \citep{hayashi_1981, martin_2012}, which define the iceline simply as a location at which the disk temperature reaches $\approx 170$\, K, here we use the opacity tables from \citet{zhu_2009} to account for the pressure dependence of the sublimation temperature.

Our calculations clearly show that for a given starting $M_d$ iceline lies at larger distance in the circumbinary disk compared to its circumstellar analogue. For example, at 1 Myr the circumbinary disk with $M_d=0.05 M_c$ has iceline at 6 AU, almost a factor of 3 further than in the circumstellar disk of the same mass. This difference arises because of the presence of strong tidal heating in the inner parts of circumbinary disks, which pushes $r_{\rm ice}$ out. In addition, viscous heating is also more significant in the inner circumbinary disk because of both the different $\FJ(r)$ distribution and the assumed lack of accretion onto the central binary, resulting in higher disk mass at late times.

The decrease of $r_{\rm ice}$ with time in the circumbinary case is driven by the viscous expansion of the disk, which lowers its characteristic $\FJ$. In circumstellar case $\FJ$ additionally decreases due to mass loss to accretion, explaining why the difference in  $r_{\rm ice}$ between the two cases grows with time. This has implications for planet formation, which are discussed in \S \ref{sect:pl_form}. 

Also, as expected, more massive disks have icelines located further out because of the increased contribution of viscous and tidal heating, both of which scale with disk surface density or $\FJ$, see equations \ref{eq:flux_nu} \& \ref{eq:ftid}).

%%%%%%%%%%%%%%%%%%%%%%%%%%%%%%%%%%%%%%%%%%%%%%%%%%%%%%%%%%%

\subsection{Limitations of our model.}  
\label{sect:limits}

%%%%%%%%%%%%%%%%%%%%%%%%%%%%%%%%%%%%%%%%%%%%%%%%%%%%%%%%%%%

We now comment on the validity of the approximations used in this work and outline its limitations (in addition to not accounting for the possibility of a dead zone, see \S \ref{sect:deadzone}). 

Our model ignores the azimuthal structure of the disk which should arise due to the rotating potential of the binary. However, in the time-averaged sense our axisymmetric model should still represent fairly well the actual disk structure. This is especially true beyond several$\times a_b$, where the time-varying non-axisymmetric disk structures become subdominant. The same is true for the variability in time and space of the irradiation flux impinging on the disk surface \citep{clanton_2013,Shadmehri,Quillen}.

Our treatment of disk thermodynamics has room for improvement. Our calculation of the midplane disk temperature ignores the possibility of the disk self-shadowing in regions where viscous and tidal heating dominate, see \S \ref{sect:shadowing}. This motivates future more detailed, higher-dimensional studies of the radiation transport in the circumbinary disks that would be able to account for disk self-shadowing.

Our assumption of equality between the temperature of the superheated grains and the midplane temperature in the irradiated case (see \S \ref{sect:spectra}) may affect the determination of $T$ when the disk becomes optically thin. The rate at which the disk viscously spreads determines evolution of many of its properties (e.g. $\FJ$, see \S \ref{sect:analytical}), and the behavior of $r_{\rm infl}$ is set by the disk temperature in the outermost, directly irradiated and often optically thin regions. This motivates more careful treatments of disk thermodynamics in the future. However, we believe that this simplification may affect our results at most at the level of tens of per cent.

We expect our results on the global disk evolution to be rather insensitive to the exact form of the binary torque density $\Lambda(r)$. However, equation (\ref{eq:ftid}) shows that heating of the innermost disk regions is directly determined by $\Lambda(r)$, for which we have adopted a simple form (\ref{eq:Lambda}). Future treatments of the circumbinary disk physics would benefit from the more accurate torque density description \citep{GR01,R02,RP12,petrovich_2012}.

Our models have typically been evolved for 5 Myr without any external mass loss from the system. In practice, disk is likely to be losing mass due to photoevaporation \citep{alexander_2012}, which will modify its structure and evolution at late times. 

%%%%%%%%%%%%%%%%%%%%%%%%%%%%%%%%%%%%%%%%%%%%%%%%%%%%%%%%%%%

\subsection{Comparison with other work.}  
\label{sect:others}

%%%%%%%%%%%%%%%%%%%%%%%%%%%%%%%%%%%%%%%%%%%%%%%%%%%%%%%%%%%

Several authors have previously addressed different aspects of the circumbinary disks and we discuss how our results fit in the context of these studies.

\citet{alexander_2012} and \cite{martin_2013} presented evolutionary circumbinary disk models based on equation (\ref{eq:evSigma}), focusing on the role of photoevaporation and dead zone, respectively. Both studies have chosen the torque density prescription in the form (\ref{eq:Lambda}) with $f=0.5$. This value is much larger than what we use in our work, resulting in wider inner cavities in these studies --- $(4-5)a_b$, which is $\approx 2$ times larger than suggested by the results of hydrodynamical simulations \citep{mcfadyen_2008,pelupessy_2013}. Because our disks extend closer to the central binary (down to $\approx 2a_b$), tidal and viscous heating drive inner disk temperature to significantly higher values in our models, up to $2000$ K as opposed to 800 K in non-accreting models models of \cite{martin_2013} with the same disk mass.

We also provide a more accurate treatment of the disk thermodynamics. For comparison, \citet{alexander_2012} have simply prescribed a fixed temperature profile in the disk. \cite{martin_2013} provide a treatment of disk thermodynamics similar to ours and account for viscous and tidal heating, albeit with a simplified opacity behavior $\kappa\propto T^{0.8}$ adopted for the whole disk. However, their prescription for irradiation heating assumes $T_{\rm irr}\propto r^{-3/4}$, which is not appropriate for the externally irradiated, flared disk. As a result, the outer regions of their disk models receive much less heating than the flared irradiated disks, which should have implications for the viscous spreading and global evolution of their disks. 

\citet{clanton_2013} explored the issue of iceline location in a circumbinary disk and found $r_{\rm ice}$ to lie rather close to the binary. However, this calculation did not account for the tidal dissipation, which as we show in this work, dominates heating in the inner disk. \citet{Shadmehri} have improved the $r_{\rm ice}$ calculation by accounting for tidal heating, and found significantly larger values of $r_{\rm ice}$. However, both of these studies did not properly account for the changes of the disk structure caused by the injection of the angular momentum at its center. Instead, they simply adopted $\Sigma(r)$ structure that would correspond to a standard constant-$\dot M$ disk with no torque at the center. As we show in our work, this assumption is not justified in general and $\Sigma(r)$ and $\FJ(r)$ distributions should be quite different in realistic circumbinary disks from their analogues in conventional constant-$\dot M$ circumstellar counterparts.

%%%%%%%%%%%%%%%%%%%%%%%%%%%%%%%%%%%%%%%%%%%%%%%%%%%%%%%%%%%%%%%%
%%%%%%%%%%%%%%%%%%%%%%%%%%%%%%%%%%%%%%%%%%%%%%%%%%%%%%%%%%%%%%%%

\section{Implications for planet formation}
\label{sect:pl_form}

%%%%%%%%%%%%%%%%%%%%%%%%%%%%%%%%%%%%%%%%%%%%%%%%%%%%%%%%%%%%%%%%

We now discuss the ramifications of our results for circumbinary planet formation. 

Suppression of accretion (even partial) by the binary torque implies that the circumbinary disk retains more mass that its circumstellar counterpart at the same moment of time. In fact, there is some observational evidence in support of this conjecture. \citet{Harris} find that circumbinary disks tend to be significantly brighter at sub-mm wavelength than their counterparts around the single stars, although one has to keep in mind that the size of their sample of circumbinary disks with detected emission is small (only four systems). Higher disk mass persisting for longer has a number of important consequences for planet formation in circumbinary disks. 

First, higher surface density of gas and solids should speed up the growth of planetesimals due to larger mass supply. For illustration, comparison of Figures \ref{fig:md0.1} and \ref{fig:md0.05_circumstellar} demonstrates that e.g. at 3 AU $\Sigma$ in a circumbinary disk is higher by $\sim 3$ compared to the circumstellar disk of the same initial mass $M_d=0.05M_c$ at $t=0.5$ Myr. This implies that in the former case planetesimal growth by coagulation would be faster by roughly the same factor.

Second, gravitational effect of a more massive disk significantly reduces eccentricity excitation by the binary gravity \citep{rafikov_2013a,SR15}. As shown by \citep{SR15}, this lowers relative speeds with which planetesimals collide, alleviating the so-called {\it fragmentation problem} for planet formation in binaries \citep{meschiari_2012,paardekooper_2012,marzari}. 

Third, the isolation mass of the cores forming in massive circumbinary disks should be larger than in their circumstellar counterparts, since it scales as $\Sigma^{3/2}$. This may allow direct formation of the planetary cores massive enough to trigger runaway gas accretion and turn into giant planets.

Our results have non-trivial implications for another issue relevant for reducing collisional velocities of planetesimals in binaries, namely the precession of the central binary. It was shown by \citet{rafikov_2013a} and \citet{SR15} that disk gravity often dominates precession of the eccentric central binary, which in turn reduces planetesimal eccentricity excitation by the non-axisymmetric component of the binary gravity. The rate of binary precession is set mainly by the disk surface density at its inner edge, which was previously estimated using simple models of passively heated (externally irradiated) disks \citep{rafikov_2013a,SR15}. However, our results show (see Figures \ref{fig:md0.1}-\ref{fig:md0.01}) that because of the tidal and viscous heating $T$ is considerably higher in the inner disk regions than equation (\ref{eq:T_irr}) would predict, implying higher viscosity and lower $\Sigma$ at the cavity edge compared to the prediction (\ref{eq:surf_irr}). This may lower the precession rate of the central binary by a factor of several compared to the existing models.

On the other hand, outward displacement of the iceline by the vigorous tidal and viscous heating may negatively affect core assembly in the inner (within few AU) regions of circumbinary disks by reducing the surface density of solids (per $\Sigma$ of gas) there. This effect may favor formation of the observed {\it Kepler} circumbinary planets in the outer regions of their parent disks (e.g. outside $\sim 5$ AU) with subsequent inward migration into their present day location, a scenario that is preferred on other grounds as well  \citep{paardekooper_2012,rafikov_2013a,SR15}.

%%%%%%%%%%%%%%%%%%%%%%%%%%%%%%%%%%%%%%%%%%%%%%%%%%%%%%%%%%%

\section{Summary.}  
\label{sect:summ}

%%%%%%%%%%%%%%%%%%%%%%%%%%%%%%%%%%%%%%%%%%%%%%%%%%%%%%%%%%%

We explored properties and viscous evolution of protoplanetary disks around stellar binaries, which are the birthplaces of circumbinary planets such as the ones detected by the {\it Kepler} mission. Our main conclusions are briefly summarized as follows.

\begin{itemize}

\item 
Circumstellar disks are in many ways different from their circumstellar analogues (\S \ref{sect:disk_prop}). Suppression of accretion onto the central object by the binary torques not only allows the disk to retain mass in the disk for longer. Tidal barrier imposed by the central binary modifies the basic character of the radial distribution of the viscous angular momentum flux $\FJ(r)$, leading to a flat $\FJ$ profile in the case of no accretion (\S \ref{sect:steady}). This has important implications for the disk structure --- circumbinary disks contain more mass and release more energy by viscous heating in their central regions than their circumstellar counterparts of the same mass. 

\item 
Very importantly, we derive a set analytical relationships for the viscous evolution of disk properties (\S \ref{sect:analytical}) that are verified and calibrated by the detailed simulations with realistic inputs (\S \ref{sect:numerical}). These relations form a basis for quantitative understanding of the role of different parameters of the system (disk mass, viscosity, etc.) on its evolution. They are  flexible and can be generalized to account for additional effects such as the non-negligible accretion onto the central binary (\S \ref{sect:someaccretion}).

\item 
Dissipation of the binary-driven density waves dominates heating of the inner disk, within 1-2 AU (\S \ref{sect:heating_terms}). This energy source (absent in disks around single stars) raises inner disk temperature, and pushes the iceline further out (to $\sim (5-10)$ AU) compared to circumstellar disks (\S \ref{sect:iceline}). Irradiation by central binary starts to control disk temperature only outside $\sim 8$ AU.

\item 
SED of a circumbinary disk is different from SED of its circumstellar counterpart of the same mass. Dissipation of the density waves gives rise to a distinctive bump in the SED around $10\mu$m that may facilitate identification of circumbinary disks when the binarity of the central source is not obvious (\S \ref{sect:spectra}).

\item 
Viscous and tidal energy release in the central region give rise to self-shadowing of the disk by its inner parts out to $\sim 20$ AU (\S \ref{sect:shadowing}).

\item 
Tidal coupling to the disk continuously removes angular momentum from the central binary, shrinking its orbit and potentially resulting in its merger into a single star (\S \ref{sect:binangloss}). 

\item 
Circumbinary disks are in many ways more favorable sites of planet formation than their analogues around single stars (\S \ref{sect:pl_form}). This is in agreement the occurrence rates of circumbinary planets inferred from statistics of {\it Kepler} systems \citep{Armstrong}.

\end{itemize}

Our study thus provides a basis for understanding the systems in which {\it Kepler} circumbinary planets were born.

\acknowledgements

Authors are grateful to Zhaohuan Zhu and Phil Armitage for useful discussions. RRR is an IBM Einstein Fellow at the IAS. He thanks Lebedev Physical Institute for hospitality during the final stages of this work. Financial support for this study has been provided by NSF via grants AST-1409524,  AST-1515763, NASA via grant 14-ATP14-0059, and The Ambrose Monell Foundation.

%%%%%%%%%%%%%%%%%%%%%%%%%%%%%%%%%%%%%%%%%%%%%%%%%%%%%%%%%%%
%%%%%%%%%%%%%%%%%%%%%%%%%%%%%%%%%%%%%%%%%%%%%%%%%%%%%%%%%%%

 \nocite{orosz_2012b}
 \nocite{rafikov_2013b}
 \nocite{lin_1979b}
 \bibliographystyle{apj}
 \bibliography{references}
 
%%%%%%%%%%%%%%%%%%%%%%%%%%%%%%%%%%%%%%%%%%%%%%%%%%%%%%%%%%%

\appendix 

%%%%%%%%%%%%%%%%%%%%%%%%%%%%%%%%%%%%%%%%%%%%%%%%%%%%%%%%%%%
%%%%%%%%%%%%%%%%%%%%%%%%%%%%%%%%%%%%%%%%%%%%%%%%%%%%%%%%%%%

\section{Numerical setup for circumbinary disk evolution.}  
\label{sect:setup}

%%%%%%%%%%%%%%%%%%%%%%%%%%%%%%%%%%%%%%%%%%%%%%%%%%%%%%%%%%%

To explore viscous evolution of a protoplanetary disk we numerically solve equation (\ref{eq:evSigma}). We first recast it as a mass conservation law with an advection term, and then use the finite-volume method to discretize it with an explicit-upwind scheme. The time step size is determined adaptively by computing the minimum time for any cell in the grid with nonzero mass to be emptied if the fluxes are held constant; then a small fraction ($0.01 \%$) of that time is used as the time step size. We use the software package FiPy \citep{guyer_2009} for the discretization and the integration of equation (\ref{eq:evSigma}). Our source code is freely available\footnote{https://github.com/garmilla/circumbinary}.

We use a logarithmically spaced radial grid that spans from $r_{in}=0.2$ AU to $r_{out}=2 \times 10^4$ AU with 150 grid points (we verified that our results are converged at this resolution). We consider a central binary with $q=1$, $M_c=M_{\odot}$ (i.e. $M_p=M_s=0.5M_\odot$), $a_b=0.2$ AU, $L_c=2\,L_{\odot}$, and three disk masses $M_d=0.01$ $M_c$, 0.05 $M_c$, and 0.1 $M_c$. The viscous $\alpha$-parameter is set to $\alpha=0.01$ in all our runs. 

Disk thermodynamics enters equation (\ref{eq:evSigma}) only via the $\alpha$-parametrization of viscosity (\ref{eq:alpha_ansatz}), and is self-consistently treated by solving equation (\ref{eq:gen_T}) at each time step.  Our temperature prescription thus allows for viscous, tidal, and irradiation heating and interpolates over optically-thick and -thin limits. 

Disk irradiation by the binary is treated in a simplified way. Namely, we assume that the incidence angle $\zeta$ is always that of a purely irradiation-dominated disk. For that reason $\zeta$ is not computed using equation (\ref{eq:zeta}) with the actual $T(r)$ setting $h(r)$. Instead, $\zeta$ is simply given by equation (\ref{eq:zeta_irr}). As the disk may become non-flared at certain radii, shadowing exterior portions of itself (which we indeed find to be the case, see \S \ref{sect:shadowing}), the irradiation may be absent in some parts of the disk. For simplicity, we neglect this self-shadowing effect and discuss the ramifications of this approximation in \S \ref{sect:shadowing}.

To estimate optical depth, we adopt the opacity fits from \citet{zhu_2009}, which assume $\kappa$ behavior in the form $\kappa = \kappa_i T^{a_i}\rho^{b_i}$ in a number of regions in $T$-$\rho$ space. Parameters $\kappa_i$, $a_i$, $b_i$ depend on the physical condition in the $i$-th opacity regime, and change as dust particles of different composition sublimate, molecules are dissociated, and different species are ionized (see \citet{zhu_2009} for details). This $\kappa$ prescription spans from electron scattering opacity at high $T$ to opacity dominated by water ice grains at low $T$.

As our initial condition we start the disk as a narrow ring of mass at some initial radius $r_0$ and let it evolve viscously. The initial radial distribution of $\Sigma$ is given by
a Gaussian ring of width $\sigma_r\ll r_0$
\ba
  \Sigma(r,t=0) \approx \frac{M_d}{(2\pi)^{3/2}r_0\sigma_{r}}
  \exp\left[-\frac{\left(r-r_0\right)^{2}}{2 \sigma_{r}^2}\right],
  \label{eq:initial-cond}
\ea
where we took $r_0=20$ AU and $\sigma_r =2$ AU. After several viscous timescales our results should not depend on this initial condition.

We use the following boundary conditions (BCs). At the outer boundary of our computation domain we set mass accretion rate to zero:
\ba
\left.
  \frac{\partial \FJ}{\partial l}\right|_{r_{out}}=\dot M(r_{out},t)=0.
  \label{eq:outer-bdy}
\ea
In practice, our spreading disks never reach the outer boundary so that this BC is well observed. At the inner boundary, in a circumstellar case, we impose a uniform (in radius) accretion rate BC via
\ba
\left.
  \frac{\partial F_J}{\partial l} 
  \right|_{r_{in}} = 
  \frac{F_J(r_{in})}{l_{in}},
  \label{eq:inner-bdy}
\ea
where $l_{in}=l(r_{in})$ is the specific angular momentum at the inner boundary. This form of BC is motivated by equation (\ref{eq:const_Mdot}). Note that we do not specify the actual value of $\dot M$, instead, we allow $\dot M(r_{in})$ to be self-consistently determined by the viscous stresses. In a circumbinary case, because of high masses of the stellar components, $\Lambda$ in the form  (\ref{eq:Lambda}) results in vanishingly small $\Sigma(r_{in})$ and $\dot M(r_{in})$, resulting in $\dot M_b=0$ (see \S \ref{sect:far_regime}), so that the inner BC is irrelevant. In real disk, $\dot M(r_{in})$ may be different from zero because of non-axisymmetric effects not captured by our 1D approach \citep{orazio_2013}. For simplicity we do not consider this possibility here.

The injection of angular momentum in the disk center by the binary is modeled as follows. For $|r-a_b| > h(r)$, we follow the prescription (\ref{eq:Lambda}), while for $|r-a_b| < h(r)$ we use $\Lambda(r) \propto r$, continuously matching at $r=a_b\pm h(r)$. Note that this introduces a discontinuity in the derivative of $\Lambda(r)$ at this point. To avoid numerical problems we remove the discontinuity by smoothing $\Lambda$. Since we are using a high mass ratio ($q=1$) and matter never gets close to the binary, the details of the smoothing are unimportant.

Moreover, as previously stated in \S \ref{sect:evolution}, the details of the tidal disk-binary coupling occurring at $r\lesssim r_\Lambda$ are irrelevant for the global disk evolution at $r\gtrsim r_\Lambda$, as long as the central angular momentum source gives rise to a proper inner BC. This is how we exploit the prescription (\ref{eq:Lambda}), not worrying about the details of $\Lambda(r)$ but simply making sure that our 1D disk profile reproduces gross features of the more detailed 2D numerical calculations of circumbinary disks. As a particular metric for comparison we use the size of the inner cavity, which \citet{mcfadyen_2008} found to be about $2a_b$. In all our calculations we tune the value of the parameter $f$ in the expression (\ref{eq:Lambda}) so that in each of our runs cavity edge is at $r_c\approx 2a_b$. This typically results in $f\approx 10^{-3}$-$2\times10^{-3}$, with specific values for each disk model indicated in Figures \ref{fig:md0.1}-\ref{fig:md0.01}.

%%%%%%%%%%%%%%%%%%%%%%%%%%%%%%%%%%%%%%%%%%%%%%%%%%%%%%%%%%%
%%%%%%%%%%%%%%%%%%%%%%%%%%%%%%%%%%%%%%%%%%%%%%%%%%%%%%%%%%%

\section{Spectral energy calculation}
\label{ap:SED}

%%%%%%%%%%%%%%%%%%%%%%%%%%%%%%%%%%%%%%%%%%%%%%%%%%%%%%%%%%%

The energy flux ${\cal F}$, emitted by a unit surface area 
element of the disk at each radius $r$, consists of several 
contributions. First, there is a (one-sided) flux due to viscous and tidal 
dissipation ${\cal F}_v$ and ${\cal F}_{\rm tid}$ given by equations (\ref{eq:flux_nu}) and (\ref{eq:ftid}). Second, part of the incoming stellar radiation is intercepted by the 
superheated dust layer, re-emitted towards the disk, absorbed 
and then lost to space. Finally, the remaining part of stellar
radiation intercepted by superheated grains is re-emitted directly
to space. The relative partition between the last two contribution 
depends on the optical depth of the disk.

The first two flux contributions are re-emitted by the disk at 
the characteristic temperature $T_{\rm e}$, which we
compute according to the following formula:
\ba
\sigma T_{\rm e}^4=\left(1+\tau^{-1}\right)\left({\cal F}_v+{\cal F}_{\rm tid}\right)+
{\cal F}_{\rm irr}.
\label{eq:T_e}
\ea 
In the optically thin limit this expression reduces to the midplane disk temperature 
given by equation (\ref{eq:T_thin}). 

Superheated dust layer emits at the temperature $T_{\rm sh}$ given by  
\citep{chiang_1997}
\ba
T_{\rm sh}(r) \approx \left[\frac{L_c}{16 \pi \sigma \epsilon(T_{\rm sh}) r^{2}}
\right]^{1/4}.
\label{eq:Td}
\ea
Here $\epsilon(T_{\rm sh})=Q_{\rm abs}(T_{\rm sh})/Q_{\rm abs}(T_\star)$ is the ratio of absorption efficiencies of small grains at $T_{\rm sh}$ and stellar temperature $T_\star$, for which we adopt the approximation of \citet{chiang_1997} $\epsilon(T)\approx T/T_\star$. 

The one-sided disk SED is then computed using the following interpolating formula, which is designed to correctly reproduce the limiting cases of optically thick and thin disk:
\ba
F_\nu=\frac{\tau}{1+\tau}B_\nu(T_{\rm e})+
\left(\frac{2+\tau}{1+\tau}\right)
\frac{{\cal F}_{\rm irr}}{\sigma T_{\rm sh}^4}B_\nu(T_{\rm sh}).
\label{eq:SED}
\ea
For simplicity we do not consider $\tau$ to be a function of frequency in this expression.
When integrated over $\nu$ this expression results in the total flux $\left({\cal F}_v+{\cal F}_{\rm tid}\right)+2{\cal F}_{\rm irr}$ for any $\tau$, in agreement with energy conservation (recall that ${\cal F}_{\rm irr}$ is a half of the stellar flux incident on a disk surface). Note that ${\cal F}_{\rm irr}/\left(\sigma T_{\rm sh}^4\right)=2\epsilon(T_{\rm sh}) \zeta$, see equations (\ref{eq:Firr}) and (\ref{eq:Td}). 

As stated in \S \ref{sect:transport}, we are ignoring the difference in values of opacity for the radiation at temperatures $T_{\rm e}$ and $T_{\rm sh}$. In the optically thick limit ($\tau\gg 1$) the amount of flux ${\cal F}_v+{\cal F}_{\rm tid}+{\cal F}_{\rm irr}$ is emitted by the disk at effective temperature $T_{\rm e}$, while another ${\cal F}_{\rm irr}$ is radiated by the superheated layer at the temperature $T_{\rm sh}$. In the limit of small optical depth ($\tau\ll 1$) disk emits only the sum of viscous and tidal energy fluxes ${\cal F}_v+{\cal F}_{\rm tid}$, while the two superheated layers (both of which are visible because the disk is transparent as $\tau\to 0$) radiate $2{\cal F}_{\rm irr}$.

Note that, as described in Appendix \ref{sect:setup}, in computing the SED we assume the grazing incidence angle of the stellar radiation to be always given by equation (\ref{eq:zeta_irr}). This assumption may not be justified in intermediate radii, which can be shadowed, as discussed in \S \ref{sect:shadowing}. Also, use of ${\cal F}_{\rm irr}$ in the form (\ref{eq:Firr}) at all distances neglects the fact that superheated dust layer becomes inefficient at absorbing stellar radiation at large radii. Neglect of this issue artificially boosts the emission produced by the superheated layer compared to the disk emission, see \S \ref{sect:spectra}.

\end{document}